\documentstyle[12pt]{article}
\begin{document}
%
\newcommand {\QGP} {quark-gluon plasma\ }
\newcommand{\beq}{\begin{equation}}
\newcommand{\eeq}{\end{equation}}
\newcommand{\beqa}{\\ \begin{minipage}{\linewidth} \begin{eqnarray}}
\newcommand{\eeqa}{\end{eqnarray} \end{minipage} \\[6mm]}
\newcommand{\barr}[1]{\begin{array}{#1}}
\newcommand{\earr}{\end{array}}
\newcommand{\no} {\nonumber}
\title{Some remarks on pion condensation
\\
in relativistic heavy ion collisions}
\author{{\bf C. Greiner, C. Gong and B. M\"uller},\\
Department of Physics, Duke University, Durham NC 27708-0305}
\date{}
\maketitle
\thispagestyle{empty}
\begin{abstract}
Recently it was pointed out that coherent or condensated states of
pions may account for the explanation of the Centauro events observed
in cosmic ray showers. We argue that an occurrence of condensed pions
requires that the system evolves far out of thermal equilibrium.
Besides an unusual charge ratio distribution we show that such a
produced state also would lead to strong isospin correlations.
\end{abstract}
\vspace*{6cm}
\parskip0mm
\begin{center}
{{Duke TH-93-53}} \hspace*{1cm} -- \hspace*{1cm} July 1993
{{hep-ph 9307336}} \hspace*{1cm} -- \hspace*{1cm} July 1993
\end{center}
\parskip4mm
\newpage
\pagenumbering{arabic}
Heavy ion collisions at high initial bombarding energies offer the possibility
to study hadronic matter at high temperatures and energy densities (up to
several GeV/fm$^{3}$). Because of their small masses pions form the most
abundant component of hadronic matter below the QCD phase transition.
In this note we address the question whether the evolution of a
thermalized hadronic system allows for a substantial number of pions to
condense into the zero-momentum state at moderate temperatures.

Rajagopal and Wilczek \cite{Ra92} recently speculated that in a
rapid quench through the second order chiral QCD phase transition
disordered chiral configurations may emerge. The scenario explicitly
assumes large deviations from thermal equilibrium. The spontaneous
growth and subsequent decay
of these configurations would give rise to large collective
fluctuations in the number of produced neutral pions compared to
charged pions, and thus could provide a mechanism explaining a
family of peculiar cosmic ray events, the Centauros \cite{La80}.
These are characterized by a large dominance of charged mesons
over electromagnetic shower particles, i.e. photons produced by
the decay of a $\pi^0$.

The large isospin fluctuations can be understood in terms of the creation of
a coherent {\it isosinglet} state of pions \cite{Ko92}. Assuming that all
pions have the same momentum, e.g. ${\bf p}=0$, the only isospin singlet state
is given by \cite{Ho71}
\beq
\label{1}
| \psi_{2N}(I=0) \,  \rangle \, = \,
\frac{1}{\sqrt{(2N+1)!}} \, \left( 2\hat{a}^{\dagger}_{+} \hat{a}^{\dagger}_{-}
\, - \, (\hat{a}^{\dagger}_{0})^{2} \right) ^{N} \, | 0 \rangle \, \, \, ,
\eeq
where the creation operators refer to pions.

The reason for (\ref{1}) being the sole singlet state that can be constructed
is the Bose symmetry, which requires
the overall wavefunction to be symmetric under particle exchange.
For $2N$ pions in the same momentum state there exist only one completely
symmetric (I=0) - wavefunction. The probability of finding $2n$ neutral pions
in such a state is given by the square
\beq
p_{2n}^{\pi^0} \, = \, \left| \langle n,N | \psi _{2N} (I=0) \, \rangle
\right| ^{2} \, \, \, ,
\eeq
where
\beq
|n,N \rangle \, = \, \frac{1}{\sqrt{(2n)!}} \frac{1}{(N-n)!} \,
(\hat{a}^{\dagger}_{0})^{2n}(\hat{a}^{\dagger}_{+} \hat{a}^{\dagger}_{-}
)^{(N-n)}
|0\rangle \, \, \, ,
\eeq
and is evaluated to be
\beq
\label{2}
p_{2n}^{\pi^0} \, = \, \left( \frac{N!}{n!} \right)^2
\frac{(2n)!}{(2N+1)!} \, 2^{2(N-n)} \, \stackrel{N,n \gg 1}{\longrightarrow} \,
\frac{1}{2\sqrt{N}\sqrt{n}} \, \propto \,
\frac{1}{\sqrt{n}} \, \, \, .
\eeq
The distribution (\ref{2}) implies that the highest probability
occurs for events with a low number of neutral pions, although
$\langle \hat{n}_{\pi^{0}} \rangle =
\langle \hat{n}_{\pi^{+}} \rangle =
\langle \hat{n}_{\pi^{-}} \rangle = \frac{2}{3}N$
is exactly satisfied by (\ref{1}).

In addition, there is also a strong correlation among the various pion pair
channels. For example, one finds
\beq
\begin{array}{rcl}
\langle \hat{n}_{\pi^{0}}\hat{n}_{\pi^{0}} \rangle & = &
\langle \, \psi _{2N} | \hat{a}^{\dagger}_0 \hat{a}_0 \hat{a}^{\dagger}_0
\hat{a}_0 | \psi _{2N} \, \rangle \\
&& \\
& = & \frac{4}{5} N^2 \, + \, \frac{8}{15} N \,
\stackrel{N \gg 1}{\longrightarrow} \frac{4}{5} N^2 \, \,  \neq \,  \,
\langle \hat{n}_{\pi^0} \rangle ^2 \, = \, \frac{4 }{9} N^2 \, \, \, .
\end{array}
\eeq
More generally, the correlations
\beq
C_{\alpha\beta} \, \equiv  \,
\frac{\langle \hat{n}_{\pi^{\alpha}} \hat{n}_{\pi^{\beta}} \rangle}
{\langle \hat{n}_{\pi^{\alpha}}
\rangle \langle \hat{n}_{\pi^{\beta }}\rangle } - 1
\eeq
are for large $N$ found to be
\beqa
\label{3}
C_{00}  & \to & + \frac{4}{5},   \nonumber \\
C_{0+}  & \to & - \frac{2}{5},    \\
C_{++}  & \to & + \frac{1}{5},   \, \, \, \mbox{\rm etc}. \nonumber
\eeqa
Hence, besides yielding a peculiar asymmetry in the pion charge
distribution, the particular isospin coupling
results in a correlation among pions
of equal charge and an anti-correlation among different pion pairs.
Such correlations would be highly unusual and hard to explain otherwise.
In intensity interferometric experiments with bosons one usually expects
{\it no} correlation for a coherent state, i.e.  $C = 0$ \cite{Gy79}.

Regardless of this speculative scenario of decaying chiral condenates,
it is useful to ask whether a condensed pion state can be formed under
circumstances where the hadronic matter never gets far out of thermal
equilibrium.  It is expected
that the pion density becomes quite large at high bombarding energies
and for heavy colliding nuclei. If a locally thermalized state is
created throughout the rapidity gap between projectile and target nuclei,
can the density become so high that some fraction of the pions are forced
into a condensate?

This would require that the pions evolve out of {\it chemical} equilibrium
up to a point, where their chemical potential equals the effective pion mass.
Indeed, a chemical potential of $\mu _{\pi} \approx 90 - 130$
MeV was introduced to explain the slopes of $p_T$-spectra observed in heavy
ion collisions which show a characteristic enhancement at low transverse
momentum \cite{Ka90,Ge90}. At presently accessible energies there is still a
non-negligible density of baryons present at midrapidity due to partial
stopping of baryons, which provides a large `reservoir' of isospin for
the pion gas. However, at future collider energies it is predicted \cite{Mu92}
that gluons will initially equilibrate and thermalize more rapidly than quarks
and antiquarks, and presumably the midrapidity region is almost completely
baryon-free. Since gluons carry no isospin, they must yield, after
hadronization, particles that are in an isospin singlet state.
For the moment let us assume that a major fraction of these pions would
form a condensate state which carries no isospin.
This would look precisely like the state (\ref{1}) constructed above and
would give rise to the same isospin fluctuations and correlations.

At temperatures between 100 and 200 MeV the pion dispersion relation
changes due to the soft interaction among pions, which is dominated by the
attractive channel of the $\rho $-meson resonance.
A simple representation for the induced self-energy
$\Sigma (p)$ of the pion is \cite{Sh91}
\beq
\label{4}
\Sigma_{\pi}(p) \, = \,
\int \frac{d^3k}{2\omega_k} \, \left(- \, {\cal M}_{\pi \pi}(k,p) \right) \,
n_{\pi} ({\bf k}) \, \, \, ,
\eeq
where ${\cal M}$ describes the forward $\pi-\pi$ scattering amplitude.
It can be evaluated using the experimental scattering data for two pions.
Assuming that its imaginary part
$\Im \{ \, \Sigma \}$ is sufficiently small, the modified
dispersion relation of the interacting pions reads
\beq
\label{5}
E({\bf p}) \, = \, \left(p^2 + m_{\pi }^2 + \Re \left\{ \, \Sigma
 (E({\bf p}),\, {\bf p})\right\} \right)
^{\frac{1}{2}} \, \, \, .
\eeq
In Fig. 1a the critical chemical potential,
$\mu_c=(m_{\pi}^2+\Re \{ \, \Sigma (\mu_c,{\bf 0})\} )^{1/2}$,
where Bose condensation sets in, is depicted as a function of temperature.
For temperatures at the lower end of the range $T\approx 100-120 $ MeV
the decrease in $\mu_c$ is small
(about $5-10$ MeV compared to the free pion mass), but $\mu_c$ decreases
quite strongly for higher temperatures. For convenience, the pion density at
$\mu_\pi = \mu_c(T)$ and at $\mu_\pi =0$ are shown in Fig. 1b.
At the critical potential the density is at least twice as large as
the density for pions in chemical equilibrium at the same temperature.
At higher $T$ the necglect of higher than two-particle interactions may not
be adequate, because the pion gas is not really dilute. However, for our
present purpose the treatment is sufficient.

The chemical potentials deduced from the observed $p_T$-spectra
\cite{Ka90,Ge90} with the assumption of a noninteracting pion gas are in
the same range as the critical chemical potentials.
Microscopically this soft mode ``excess'' stems from the final state
(elastic) collisions increased by the Bose enhancement factor $(1+f)$
\cite{We92}.
One may expect that
this would present an opportunity for the onset of condensation. We will
argue now, that this is not the case.

If the hadronization out of a deconfined \QGP occurs rather slowly
at some critical temperature $T_c\approx 150 - 220$ MeV,
the pions should be in chemical equilibrium with their environment,
in particular with all the mesonic resonances. If the hadronic phase
is formed at an initial temperature larger than $T\approx 170$ MeV,
the inelastic reactions, which change the pion number, enforce
$\mu_{\pi}=0$ \cite{Be92}. Below this temperature the scattering rates
for the inelastic processes are getting small, so that during the subsequent
cooling of the hadronic phase the pions may drop out of
chemical equilibrium. Thermal equilibrium, of course, is maintained
down to a decoupling temperature
of $T\approx 100 - 120$ MeV due to the large elastic collisions
preventing the system from an early freeze-out. In the absence
of inelastic reactions the number of pions must stay constant ($dN_\pi=0$),
because pions cannot be absorbed or annihilated.
Because of overall energy conservation ($dE+pdV=0$) the
entropy also stays constant:
\beq
dS=-\frac{\mu_\pi}{T} \, dN_{\pi} = 0.
\eeq
Thus the ratio $s_{\pi} / n_{\pi}$ is conserved below $T\approx 170$ MeV.

In Figs. 2 the entropy per pion $s_{\pi}/n_{\pi}$ is shown as function of $T$.
Fig. 2a is calculated for the interacting pion gas with the modified
dispersion relation (\ref{5}), whereas Fig. 2b is for noninteracting pions.
Both figures show the evolution of this ratio with temperature for several
values of the chemical potential $\mu_{\pi}$. The most important of the
figures is that, along the line following the critical chemical potential,
$s_{\pi}/n_{\pi}$ {\em decreases} with falling temperature.

We can now adopt the reasoning of the work by Bebie et al \cite{Be92}.
If, for example, the pions lose their chemical equilibrium at a temperature
$T\approx 180 $ MeV, the required constant ratio of  $s/n$ drives the
system towards nonvanishing $\mu_{\pi}>0$ during the cooling process.
At a freeze-out temperature around 100
MeV the system ends up with a chemical potential $\mu_\pi = 60-80$ MeV
independent of interactions. The incorporation of mesonic
resonances, like the $\rho$ and $\omega$, leads to a somewhat higher
chemical potential $\mu_{\pi} \approx 100$ MeV at the decoupling temperature.
Yet the line along the critical potential is not seriously affected and
still drops for decreasing temperature.
Because of this peculiar behaviour of $s/n$ at $\mu_\pi = \mu_c$, a pion gas
in thermal equilibrium can never reach the onset of Bose condensation.
This could only happen if additional pions were produced near the
decoupling temperature, but we know of no mechanism which would do this.

The phase transition of the \QGP phase may occur at some critical temperature
$T_c<170$ MeV. The pions would then be produced in chemical equilibrium,
if the transition is sufficiently slow \cite{Be88},
but later would rather quickly
drop out of chemical equilibrium during the expansion. Again a chemical
potential would be built up which, according to our argument,
should be smaller than that obtained for pions from an initially hotter
hadronic phase.

The only possible mechanism producing a supercritical potential, and
hence a pion condensate, is during the phase transition.
This can only be achieved if the
transition itself is fast compared to the typical inverse
scattering rates for the inelastic channels, which are of the order of a few
fm/$c$. Then the initial hadronic phase may be far from a
chemically (and thermally) equilibrated state
and may have a sufficiently high pion density. If an initial fraction
of pions exists in a condensed state, the condensate could conceivably
survive the further evolution of the system.
Such a scenario of an abrupt phase transition by necessity corresponds to
a process of ``quenching'' as assumed by Rajagopal and Wilczek \cite{Ra92}.

Recently \cite{Pr92} it was argued that a very dense (and thermal)
pion source affects the single particle distribution,
the two-particle momentum correlation function $C(k)$ and the
isospin distribution. However, the densities required for such
a ``paser'' \cite{Lam} producing significant effects, turns out to be
at least some 1/fm$^3$, already corresponding to a certain fraction of
Bose condensed pions. Especially the density necessary for demonstrating
the laser feedback on the emission of the three different isospin states
and thus the broadening in the isospin distribution was
75 pions contained in a sphere with a radius of about 0.75 fm$^3$, which
corresponds to $n_{\pi}=50$ fm$^3$! It is hard to imagine a
pion gas at such densities.

As a final remark, refering to the consequences outlined in (\ref{2},\ref{3}),
it
would be interesting to investigate the isospin distribution and the
correlations of the pions in a thermal isosinglet state, with a chemical
potential of about 100 MeV at the freeze-out temperature.
Then, of course, many more $(I=0)$ -states would contribute to the density
matrix. Only an explicit calculation can show, whether similar
correlations are present in this case or whether these are
washed out by the mixture of participant states.

Concludingly, we have explained that the pions should never condense,
if the hadronic matter stays in thermal equilibrium throughout the evolution
and cooling process, although the pions should loose their chemical
equilibrium at lower temperatures. Only for the case if the hadronization
out of the \QGP is rather spontaneous (on the timescale of inelastic
collisions), a fraction of pions could settle in a Bose condensated state
and this state may survive until decoupling. In addition to the unusual
distribution in the charged versus neutral particle ratio the existence of
such a coherent state can also be recognized by the measurement of the various
isospin correlations.
\\[15mm]
{\bf Acknowledgements:}
\\[5mm]
One of us (C.G.) wants to thank the Alexander von Humboldt Stiftung
for its support with a Feodor Lynen scholarship. This work was supported
in part by the U.S. Department of Energy (grant DE-FG05-90ER40592).
\vspace*{15mm}
\parskip0mm
\par
\newpage
\newpage
\parindent 0mm
{\Large {\bf Figure captions}}:
\\[2cm]
{\bf Figure 1}:  \\
The critical chemical potential $\mu_c(T)$ is shown as
a function of temperature (a). Two pion interactions by the
attractive channel of the $\rho$-meson are taken into account.
In addition, the pion number densities for some chemical potentials
$\mu_{\pi}$ are also depicted as a function of $T$ (b).
\\[1cm]
{\bf Figure 2}: \\
The entropy per pion $s_{\pi}/n_{\pi}$ is plotted as function of temperature
$T$ for several fixed chemical potentials $\mu_\pi$. (a) represents the
case including the pion interactions, (b) the case for a noninteracting
pion gas (as a guide, $s/n$=3.6 for a massless and noninteracting
Bose gas).
\\
Note, that the ratio $s/n$ decreases for smaller temperatures
at the critical chemical potential $\mu_c(T)$ in both cases.
%
%


 20 dict begin
72 300 div dup scale
1 setlinejoin 0 setlinecap
/Helvetica findfont 55 scalefont setfont
/B { stroke newpath } def /F { moveto 0 setlinecap} def
/C { CS M 1 1 3 { pop 3 1 roll 255 div } for SET_COLOUR } def
/CS { currentpoint stroke } def
/CF { currentpoint fill } def
/L { lineto } def /M { moveto } def
/P { moveto 0 1 rlineto stroke } def
/T { 1 setlinecap show } def
errordict /nocurrentpoint { pop 0 0 M currentpoint } put
/SET_COLOUR { pop pop pop } def
 80 600 translate
gsave
CS [] 0 setdash M
CS M 2 setlinewidth
/P { moveto 0 2.05 rlineto stroke } def
 0 0 0 C
CS [] 0 setdash M
255 255 M 2261 255 L
346 255 M 346 280 L
486 255 M 486 280 L
626 255 M 626 306 L
767 255 M 767 280 L
907 255 M 907 280 L
1047 255 M 1047 280 L
1187 255 M 1187 280 L
1328 255 M 1328 306 L
1468 255 M 1468 280 L
1608 255 M 1608 280 L
1749 255 M 1749 280 L
1889 255 M 1889 280 L
2029 255 M 2029 306 L
2169 255 M 2169 280 L
554 182 M 569 223 M 574 226 L
581 233 L
581 182 L
578 230 M 578 182 L
569 182 M 590 182 L
624 233 M 617 230 L
612 223 L
610 211 L
610 204 L
612 192 L
617 185 L
624 182 L
629 182 L
636 185 L
641 192 L
643 204 L
643 211 L
641 223 L
636 230 L
629 233 L
624 233 L
619 230 L
617 228 L
614 223 L
612 211 L
612 204 L
614 192 L
617 187 L
619 185 L
624 182 L
629 182 M 634 185 L
636 187 L
639 192 L
641 204 L
641 211 L
639 223 L
636 228 L
634 230 L
629 233 L
672 233 M 665 230 L
660 223 L
658 211 L
658 204 L
660 192 L
665 185 L
672 182 L
677 182 L
684 185 L
689 192 L
691 204 L
691 211 L
689 223 L
684 230 L
677 233 L
672 233 L
667 230 L
665 228 L
663 223 L
660 211 L
660 204 L
663 192 L
665 187 L
667 185 L
672 182 L
677 182 M 682 185 L
684 187 L
687 192 L
689 204 L
689 211 L
687 223 L
684 228 L
682 230 L
677 233 L
1256 182 M 1270 223 M 1275 226 L
1282 233 L
1282 182 L
1280 230 M 1280 182 L
1270 182 M 1292 182 L
1316 233 M 1311 209 L
1316 214 L
1323 216 L
1330 216 L
1337 214 L
1342 209 L
1345 202 L
1345 197 L
1342 190 L
1337 185 L
1330 182 L
1323 182 L
1316 185 L
1313 187 L
1311 192 L
1311 194 L
1313 197 L
1316 194 L
1313 192 L
1330 216 M 1335 214 L
1340 209 L
1342 202 L
1342 197 L
1340 190 L
1335 185 L
1330 182 L
1316 233 M 1340 233 L
1316 230 M 1328 230 L
1340 233 L
1373 233 M 1366 230 L
1361 223 L
1359 211 L
1359 204 L
1361 192 L
1366 185 L
1373 182 L
1378 182 L
1386 185 L
1390 192 L
1393 204 L
1393 211 L
1390 223 L
1386 230 L
1378 233 L
1373 233 L
1369 230 L
1366 228 L
1364 223 L
1361 211 L
1361 204 L
1364 192 L
1366 187 L
1369 185 L
1373 182 L
1378 182 M 1383 185 L
1386 187 L
1388 192 L
1390 204 L
1390 211 L
1388 223 L
1386 228 L
1383 230 L
1378 233 L
1957 182 M 1966 223 M 1969 221 L
1966 218 L
1964 221 L
1964 223 L
1966 228 L
1969 230 L
1976 233 L
1986 233 L
1993 230 L
1995 228 L
1998 223 L
1998 218 L
1995 214 L
1988 209 L
1976 204 L
1971 202 L
1966 197 L
1964 190 L
1964 182 L
1986 233 M 1991 230 L
1993 228 L
1995 223 L
1995 218 L
1993 214 L
1986 209 L
1976 204 L
1964 187 M 1966 190 L
1971 190 L
1983 185 L
1991 185 L
1995 187 L
1998 190 L
1971 190 M 1983 182 L
1993 182 L
1995 185 L
1998 190 L
1998 194 L
2027 233 M 2019 230 L
2015 223 L
2012 211 L
2012 204 L
2015 192 L
2019 185 L
2027 182 L
2031 182 L
2039 185 L
2043 192 L
2046 204 L
2046 211 L
2043 223 L
2039 230 L
2031 233 L
2027 233 L
2022 230 L
2019 228 L
2017 223 L
2015 211 L
2015 204 L
2017 192 L
2019 187 L
2022 185 L
2027 182 L
2031 182 M 2036 185 L
2039 187 L
2041 192 L
2043 204 L
2043 211 L
2041 223 L
2039 228 L
2036 230 L
2031 233 L
2075 233 M 2068 230 L
2063 223 L
2060 211 L
2060 204 L
2063 192 L
2068 185 L
2075 182 L
2080 182 L
2087 185 L
2092 192 L
2094 204 L
2094 211 L
2092 223 L
2087 230 L
2080 233 L
2075 233 L
2070 230 L
2068 228 L
2065 223 L
2063 211 L
2063 204 L
2065 192 L
2068 187 L
2070 185 L
2075 182 L
2080 182 M 2084 185 L
2087 187 L
2089 192 L
2092 204 L
2092 211 L
2089 223 L
2087 228 L
2084 230 L
2080 233 L
255 2261 M 2261 2261 L
346 2261 M 346 2235 L
486 2261 M 486 2235 L
626 2261 M 626 2209 L
767 2261 M 767 2235 L
907 2261 M 907 2235 L
1047 2261 M 1047 2235 L
1187 2261 M 1187 2235 L
1328 2261 M 1328 2209 L
1468 2261 M 1468 2235 L
1608 2261 M 1608 2235 L
1749 2261 M 1749 2235 L
1889 2261 M 1889 2235 L
2029 2261 M 2029 2209 L
2169 2261 M 2169 2235 L
255 255 M 255 2261 L
255 255 M 306 255 L
255 398 M 280 398 L
255 541 M 280 541 L
255 685 M 280 685 L
255 828 M 280 828 L
255 971 M 306 971 L
255 1114 M 280 1114 L
255 1258 M 280 1258 L
255 1401 M 280 1401 L
255 1544 M 280 1544 L
255 1687 M 306 1687 L
255 1831 M 280 1831 L
255 1974 M 280 1974 L
255 2117 M 280 2117 L
255 2261 M 280 2261 L
185 229 M 206 280 M 199 278 L
194 270 L
192 258 L
192 251 L
194 239 L
199 232 L
206 229 L
211 229 L
CS M
218 232 L
223 239 L
226 251 L
226 258 L
223 270 L
218 278 L
211 280 L
206 280 L
202 278 L
199 275 L
197 270 L
194 258 L
194 251 L
197 239 L
199 234 L
202 232 L
206 229 L
211 229 M 216 232 L
218 234 L
221 239 L
223 251 L
223 258 L
221 270 L
218 275 L
216 278 L
211 280 L
137 946 M 149 996 M 144 972 L
149 977 L
156 979 L
163 979 L
170 977 L
175 972 L
178 965 L
178 960 L
175 953 L
170 948 L
163 946 L
156 946 L
149 948 L
146 951 L
144 955 L
144 958 L
146 960 L
149 958 L
146 955 L
163 979 M 168 977 L
173 972 L
175 965 L
175 960 L
173 953 L
168 948 L
163 946 L
149 996 M 173 996 L
149 994 M 161 994 L
173 996 L
206 996 M 199 994 L
194 987 L
192 975 L
192 967 L
194 955 L
199 948 L
206 946 L
211 946 L
218 948 L
223 955 L
226 967 L
226 975 L
223 987 L
218 994 L
211 996 L
206 996 L
202 994 L
199 992 L
197 987 L
194 975 L
194 967 L
197 955 L
199 951 L
202 948 L
206 946 L
211 946 M 216 948 L
218 951 L
221 955 L
223 967 L
223 975 L
221 987 L
218 992 L
216 994 L
211 996 L
88 1662 M 103 1703 M 108 1705 L
115 1713 L
115 1662 L
113 1710 M 113 1662 L
103 1662 M 125 1662 L
158 1713 M 151 1710 L
146 1703 L
144 1691 L
144 1684 L
146 1672 L
151 1665 L
158 1662 L
163 1662 L
170 1665 L
175 1672 L
178 1684 L
178 1691 L
175 1703 L
170 1710 L
163 1713 L
158 1713 L
153 1710 L
151 1708 L
149 1703 L
146 1691 L
146 1684 L
149 1672 L
151 1667 L
153 1665 L
158 1662 L
163 1662 M 168 1665 L
170 1667 L
173 1672 L
175 1684 L
175 1691 L
173 1703 L
170 1708 L
168 1710 L
163 1713 L
206 1713 M 199 1710 L
194 1703 L
192 1691 L
192 1684 L
194 1672 L
199 1665 L
206 1662 L
211 1662 L
218 1665 L
223 1672 L
226 1684 L
226 1691 L
223 1703 L
218 1710 L
211 1713 L
206 1713 L
202 1710 L
199 1708 L
197 1703 L
194 1691 L
194 1684 L
197 1672 L
199 1667 L
202 1665 L
206 1662 L
211 1662 M 216 1665 L
218 1667 L
221 1672 L
223 1684 L
223 1691 L
221 1703 L
218 1708 L
216 1710 L
211 1713 L
2261 255 M 2261 2261 L
2261 255 M 2209 255 L
2261 398 M 2235 398 L
2261 541 M 2235 541 L
2261 685 M 2235 685 L
2261 828 M 2235 828 L
2261 971 M 2209 971 L
2261 1114 M 2235 1114 L
2261 1258 M 2235 1258 L
2261 1401 M 2235 1401 L
2261 1544 M 2235 1544 L
2261 1687 M 2209 1687 L
2261 1831 M 2235 1831 L
2261 1974 M 2235 1974 L
2261 2117 M 2235 2117 L
2261 2261 M 2235 2261 L
CS [] 0 setdash M
346 2205 M B
356 2195 M 356 2215 L
336 2215 L
336 2196 L
356 2196 L
CF M
486 2184 M B
496 2174 M 496 2194 L
476 2194 L
476 2174 L
496 2174 L
CF M
626 2157 M B
636 2147 M 636 2167 L
617 2167 L
617 2147 L
636 2147 L
CF M
767 2125 M B
777 2115 M 777 2134 L
757 2134 L
757 2115 L
777 2115 L
CF M
907 2086 M B
917 2076 M 917 2095 L
897 2095 L
897 2076 L
917 2076 L
CF M
1047 2039 M B
1057 2029 M 1057 2049 L
1037 2049 L
1037 2030 L
1057 2030 L
CF M
1187 1985 M B
1197 1975 M 1197 1995 L
1178 1995 L
1178 1975 L
1197 1975 L
CF M
1328 1922 M B
1338 1912 M 1338 1932 L
1318 1932 L
1318 1912 L
1338 1912 L
CF M
1468 1849 M B
1478 1839 M 1478 1859 L
1458 1859 L
1458 1839 L
1478 1839 L
CF M
1608 1765 M B
1618 1755 M 1618 1775 L
1598 1775 L
1598 1755 L
1618 1755 L
CF M
1749 1669 M B
1758 1659 M 1758 1679 L
1739 1679 L
1739 1659 L
1758 1659 L
CF M
1889 1559 M B
1899 1549 M 1899 1569 L
1879 1569 L
1879 1549 L
1899 1549 L
CF M
2029 1433 M B
2039 1423 M 2039 1443 L
2019 1443 L
2019 1423 L
2039 1423 L
CF M
2169 1289 M B
2179 1279 M 2179 1299 L
2159 1299 L
2159 1279 L
2179 1279 L
CF M
346 2205 M 486 2184 L
626 2157 L
767 2125 L
907 2086 L
1047 2039 L
1187 1985 L
1328 1922 L
1468 1849 L
1608 1765 L
1749 1669 L
1889 1559 L
2029 1433 L
2169 1289 L
CS [] 0 setdash M
1105 100 M 1126 151 M 1126 100 L
1129 151 M 1129 100 L
1112 151 M 1110 136 L
1110 151 L
1146 151 L
1146 136 L
1143 151 L
1119 100 M 1136 100 L
1216 160 M 1211 156 L
1206 148 L
1201 139 L
1199 127 L
1199 117 L
1201 105 L
1206 95 L
1211 88 L
1216 83 L
1211 156 M 1206 146 L
1203 139 L
1201 127 L
1201 117 L
1203 105 L
1206 98 L
1211 88 L
1235 151 M 1235 100 L
1237 151 M 1252 108 L
1235 151 M 1252 100 L
1268 151 M 1252 100 L
1268 151 M 1268 100 L
1271 151 M 1271 100 L
1228 151 M 1237 151 L
1268 151 M 1278 151 L
1228 100 M 1242 100 L
1261 100 M 1278 100 L
1293 120 M 1321 120 L
1321 124 L
1319 129 L
1317 132 L
1312 134 L
1305 134 L
1297 132 L
1293 127 L
1290 120 L
1290 115 L
1293 108 L
1297 103 L
1305 100 L
1309 100 L
1317 103 L
1321 108 L
1319 120 M 1319 127 L
1317 132 L
1305 134 M 1300 132 L
1295 127 L
1293 120 L
1293 115 L
1295 108 L
1300 103 L
1305 100 L
1336 151 M 1353 100 L
1338 151 M 1353 108 L
1370 151 M 1353 100 L
1331 151 M 1345 151 L
1360 151 M 1374 151 L
1384 160 M 1389 156 L
1394 148 L
1398 139 L
1401 127 L
1401 117 L
1398 105 L
1394 95 L
1389 88 L
1384 83 L
1389 156 M 1394 146 L
1396 139 L
1398 127 L
1398 117 L
1396 105 L
1394 98 L
1389 88 L
CS [] 0 setdash M
CS [] 0 setdash M
67 1119 M 33 1136 M 84 1122 L
33 1138 M 84 1124 L
40 1136 M 55 1134 L
62 1134 L
67 1138 L
67 1143 L
65 1148 L
60 1153 L
53 1158 L
33 1163 M 60 1155 L
65 1155 L
67 1158 L
67 1165 L
62 1170 L
57 1172 L
33 1165 M 60 1158 L
65 1158 L
67 1160 L
7 1201 M 12 1196 L
19 1191 L
28 1187 L
40 1184 L
50 1184 L
62 1187 L
72 1191 L
79 1196 L
84 1201 L
12 1196 M 21 1191 L
28 1189 L
40 1187 L
50 1187 L
62 1189 L
69 1191 L
79 1196 L
16 1220 M 67 1220 L
16 1223 M 60 1237 L
16 1220 M 67 1237 L
16 1254 M 67 1237 L
16 1254 M 67 1254 L
16 1256 M 67 1256 L
16 1213 M 16 1223 L
16 1254 M 16 1264 L
67 1213 M 67 1228 L
67 1247 M 67 1264 L
48 1278 M 48 1307 L
43 1307 L
38 1305 L
36 1302 L
33 1297 L
33 1290 L
36 1283 L
40 1278 L
48 1276 L
53 1276 L
60 1278 L
65 1283 L
67 1290 L
67 1295 L
65 1302 L
60 1307 L
48 1305 M 40 1305 L
36 1302 L
33 1290 M 36 1285 L
40 1280 L
48 1278 L
53 1278 L
60 1280 L
65 1285 L
67 1290 L
16 1321 M 67 1338 L
16 1324 M 60 1338 L
16 1355 M 67 1338 L
16 1317 M 16 1331 L
16 1345 M 16 1360 L
7 1370 M 12 1374 L
19 1379 L
28 1384 L
40 1386 L
50 1386 L
62 1384 L
72 1379 L
79 1374 L
84 1370 L
12 1374 M 21 1379 L
28 1382 L
40 1384 L
50 1384 L
62 1382 L
69 1379 L
79 1374 L
CS [] 0 setdash M
stroke
grestore
showpage
end

 20 dict begin
72 300 div dup scale
1 setlinejoin 0 setlinecap
/Helvetica findfont 55 scalefont setfont
/B { stroke newpath } def /F { moveto 0 setlinecap} def
/C { CS M 1 1 3 { pop 3 1 roll 255 div } for SET_COLOUR } def
/CS { currentpoint stroke } def
/CF { currentpoint fill } def
/L { lineto } def /M { moveto } def
/P { moveto 0 1 rlineto stroke } def
/T { 1 setlinecap show } def
errordict /nocurrentpoint { pop 0 0 M currentpoint } put
/SET_COLOUR { pop pop pop } def
 80 600 translate
gsave
CS [6 12] 0 setdash M
CS M 2 setlinewidth
/P { moveto 0 2.05 rlineto stroke } def
 0 0 0 C
CS [] 0 setdash M
255 255 M 2261 255 L
346 255 M 346 275 L
486 255 M 486 275 L
626 255 M 626 296 L
767 255 M 767 275 L
907 255 M 907 275 L
1047 255 M 1047 275 L
1187 255 M 1187 275 L
1328 255 M 1328 296 L
1468 255 M 1468 275 L
1608 255 M 1608 275 L
1749 255 M 1749 275 L
1889 255 M 1889 275 L
2029 255 M 2029 296 L
2169 255 M 2169 275 L
569 197 M 580 230 M 584 232 L
590 237 L
590 197 L
588 235 M 588 197 L
580 197 M 598 197 L
624 237 M 619 235 L
615 230 L
613 220 L
613 214 L
615 205 L
619 199 L
624 197 L
628 197 L
634 199 L
638 205 L
640 214 L
640 220 L
638 230 L
634 235 L
628 237 L
624 237 L
621 235 L
619 233 L
617 230 L
615 220 L
615 214 L
617 205 L
619 201 L
621 199 L
624 197 L
628 197 M 632 199 L
634 201 L
636 205 L
638 214 L
638 220 L
636 230 L
634 233 L
632 235 L
628 237 L
663 237 M 657 235 L
653 230 L
651 220 L
651 214 L
653 205 L
657 199 L
663 197 L
667 197 L
673 199 L
677 205 L
678 214 L
678 220 L
677 230 L
673 235 L
667 237 L
663 237 L
659 235 L
657 233 L
655 230 L
653 220 L
653 214 L
655 205 L
657 201 L
659 199 L
663 197 L
667 197 M 671 199 L
673 201 L
675 205 L
677 214 L
677 220 L
675 230 L
673 233 L
671 235 L
667 237 L
1270 197 M 1282 230 M 1285 232 L
1291 237 L
1291 197 L
1289 235 M 1289 197 L
1282 197 M 1299 197 L
1318 237 M 1314 218 L
1318 222 L
1324 224 L
1330 224 L
1335 222 L
1339 218 L
1341 212 L
1341 208 L
1339 203 L
1335 199 L
1330 197 L
1324 197 L
1318 199 L
1316 201 L
1314 205 L
1314 206 L
1316 208 L
1318 206 L
1316 205 L
1330 224 M 1334 222 L
1337 218 L
1339 212 L
1339 208 L
1337 203 L
1334 199 L
1330 197 L
1318 237 M 1337 237 L
1318 235 M 1328 235 L
1337 237 L
1364 237 M 1359 235 L
1355 230 L
1353 220 L
1353 214 L
1355 205 L
1359 199 L
1364 197 L
1368 197 L
1374 199 L
1378 205 L
1380 214 L
1380 220 L
1378 230 L
1374 235 L
1368 237 L
1364 237 L
1360 235 L
1359 233 L
1357 230 L
1355 220 L
1355 214 L
1357 205 L
1359 201 L
1360 199 L
1364 197 L
1368 197 M 1372 199 L
1374 201 L
1376 205 L
1378 214 L
1378 220 L
1376 230 L
1374 233 L
1372 235 L
1368 237 L
1971 197 M 1979 230 M 1981 228 L
1979 226 L
1977 228 L
1977 230 L
1979 233 L
1981 235 L
1987 237 L
1994 237 L
2000 235 L
2002 233 L
2004 230 L
2004 226 L
2002 222 L
1996 218 L
1987 214 L
1983 212 L
1979 208 L
1977 203 L
1977 197 L
1994 237 M 1998 235 L
2000 233 L
2002 230 L
2002 226 L
2000 222 L
1994 218 L
1987 214 L
1977 201 M 1979 203 L
1983 203 L
1992 199 L
1998 199 L
2002 201 L
2004 203 L
1983 203 M 1992 197 L
2000 197 L
2002 199 L
2004 203 L
2004 206 L
2027 237 M 2021 235 L
2017 230 L
2016 220 L
2016 214 L
2017 205 L
2021 199 L
2027 197 L
2031 197 L
2037 199 L
2041 205 L
2042 214 L
2042 220 L
2041 230 L
2037 235 L
2031 237 L
2027 237 L
2023 235 L
2021 233 L
2019 230 L
2017 220 L
2017 214 L
2019 205 L
2021 201 L
2023 199 L
2027 197 L
2031 197 M 2035 199 L
2037 201 L
2039 205 L
2041 214 L
2041 220 L
2039 230 L
2037 233 L
2035 235 L
2031 237 L
2066 237 M 2060 235 L
2056 230 L
2054 220 L
2054 214 L
2056 205 L
2060 199 L
2066 197 L
2069 197 L
2075 199 L
2079 205 L
2081 214 L
2081 220 L
2079 230 L
2075 235 L
2069 237 L
2066 237 L
2062 235 L
2060 233 L
2058 230 L
2056 220 L
2056 214 L
2058 205 L
2060 201 L
2062 199 L
2066 197 L
2069 197 M 2073 199 L
2075 201 L
2077 205 L
2079 214 L
2079 220 L
2077 230 L
2075 233 L
2073 235 L
2069 237 L
255 2261 M 2261 2261 L
346 2261 M 346 2240 L
486 2261 M 486 2240 L
626 2261 M 626 2219 L
767 2261 M 767 2240 L
907 2261 M 907 2240 L
1047 2261 M 1047 2240 L
1187 2261 M 1187 2240 L
1328 2261 M 1328 2219 L
1468 2261 M 1468 2240 L
1608 2261 M 1608 2240 L
1749 2261 M 1749 2240 L
1889 2261 M 1889 2240 L
2029 2261 M 2029 2219 L
2169 2261 M 2169 2240 L
255 255 M 255 2261 L
255 255 M 296 255 L
255 355 M 275 355 L
255 455 M 275 455 L
255 556 M 275 556 L
255 656 M 296 656 L
255 756 M 275 756 L
255 857 M 275 857 L
255 957 M 275 957 L
255 1057 M 296 1057 L
255 1157 M 275 1157 L
255 1258 M 275 1258 L
255 1358 M 275 1358 L
255 1458 M 296 1458 L
255 1559 M 275 1559 L
255 1659 M 275 1659 L
255 1759 M 275 1759 L
255 1859 M 296 1859 L
255 1960 M 275 1960 L
255 2060 M 275 2060 L
255 2160 M 275 2160 L
255 2261 M 296 2261 L
199 235 M 216 275 M 210 273 L
206 267 L
CS M
205 258 L
205 252 L
206 242 L
210 236 L
216 235 L
220 235 L
226 236 L
230 242 L
231 252 L
231 258 L
230 267 L
226 273 L
220 275 L
216 275 L
212 273 L
210 271 L
208 267 L
206 258 L
206 252 L
208 242 L
210 238 L
212 236 L
216 235 L
220 235 M 224 236 L
226 238 L
228 242 L
230 252 L
230 258 L
228 267 L
226 271 L
224 273 L
220 275 L
141 636 M 158 676 M 153 674 L
149 668 L
147 659 L
147 653 L
149 643 L
153 638 L
158 636 L
162 636 L
168 638 L
172 643 L
174 653 L
174 659 L
172 668 L
168 674 L
162 676 L
158 676 L
154 674 L
153 672 L
151 668 L
149 659 L
149 653 L
151 643 L
153 640 L
154 638 L
158 636 L
162 636 M 166 638 L
168 640 L
170 643 L
172 653 L
172 659 L
170 668 L
168 672 L
166 674 L
162 676 L
189 640 M 187 638 L
189 636 L
191 638 L
189 640 L
206 668 M 208 667 L
206 665 L
205 667 L
205 668 L
206 672 L
208 674 L
214 676 L
222 676 L
228 674 L
230 672 L
231 668 L
231 665 L
230 661 L
224 657 L
214 653 L
210 651 L
206 647 L
205 641 L
205 636 L
222 676 M 226 674 L
228 672 L
230 668 L
230 665 L
228 661 L
222 657 L
214 653 L
205 640 M 206 641 L
210 641 L
220 638 L
226 638 L
230 640 L
231 641 L
210 641 M 220 636 L
228 636 L
230 638 L
231 641 L
231 645 L
141 1037 M 158 1077 M 153 1075 L
149 1070 L
147 1060 L
147 1054 L
149 1045 L
153 1039 L
158 1037 L
162 1037 L
168 1039 L
172 1045 L
174 1054 L
174 1060 L
172 1070 L
168 1075 L
162 1077 L
158 1077 L
154 1075 L
153 1073 L
151 1070 L
149 1060 L
149 1054 L
151 1045 L
153 1041 L
154 1039 L
158 1037 L
162 1037 M 166 1039 L
168 1041 L
170 1045 L
172 1054 L
172 1060 L
170 1070 L
168 1073 L
166 1075 L
162 1077 L
189 1041 M 187 1039 L
189 1037 L
191 1039 L
189 1041 L
222 1073 M 222 1037 L
224 1077 M 224 1037 L
224 1077 M 203 1048 L
233 1048 L
216 1037 M 230 1037 L
141 1438 M 158 1478 M 153 1476 L
149 1471 L
147 1461 L
147 1455 L
149 1446 L
153 1440 L
158 1438 L
162 1438 L
168 1440 L
172 1446 L
174 1455 L
174 1461 L
172 1471 L
168 1476 L
162 1478 L
158 1478 L
154 1476 L
153 1475 L
151 1471 L
149 1461 L
149 1455 L
151 1446 L
153 1442 L
154 1440 L
158 1438 L
162 1438 M 166 1440 L
168 1442 L
170 1446 L
172 1455 L
172 1461 L
170 1471 L
168 1475 L
166 1476 L
162 1478 L
189 1442 M 187 1440 L
189 1438 L
191 1440 L
189 1442 L
228 1473 M 226 1471 L
228 1469 L
230 1471 L
230 1473 L
228 1476 L
224 1478 L
218 1478 L
212 1476 L
208 1473 L
206 1469 L
205 1461 L
205 1449 L
206 1444 L
210 1440 L
216 1438 L
220 1438 L
226 1440 L
230 1444 L
231 1449 L
231 1451 L
230 1457 L
226 1461 L
220 1463 L
218 1463 L
212 1461 L
208 1457 L
206 1451 L
218 1478 M 214 1476 L
210 1473 L
208 1469 L
206 1461 L
206 1449 L
208 1444 L
212 1440 L
216 1438 L
220 1438 M 224 1440 L
228 1444 L
230 1449 L
230 1451 L
228 1457 L
224 1461 L
220 1463 L
141 1839 M 158 1880 M 153 1878 L
149 1872 L
147 1862 L
147 1856 L
149 1847 L
153 1841 L
158 1839 L
162 1839 L
168 1841 L
172 1847 L
174 1856 L
174 1862 L
172 1872 L
168 1878 L
162 1880 L
158 1880 L
154 1878 L
153 1876 L
151 1872 L
149 1862 L
149 1856 L
151 1847 L
153 1843 L
154 1841 L
158 1839 L
162 1839 M 166 1841 L
168 1843 L
170 1847 L
172 1856 L
172 1862 L
170 1872 L
168 1876 L
166 1878 L
162 1880 L
189 1843 M 187 1841 L
189 1839 L
191 1841 L
189 1843 L
214 1880 M 208 1878 L
206 1874 L
206 1868 L
208 1864 L
214 1862 L
222 1862 L
228 1864 L
230 1868 L
230 1874 L
228 1878 L
222 1880 L
214 1880 L
210 1878 L
208 1874 L
208 1868 L
210 1864 L
214 1862 L
222 1862 M 226 1864 L
228 1868 L
228 1874 L
226 1878 L
222 1880 L
214 1862 M 208 1860 L
206 1858 L
205 1855 L
205 1847 L
206 1843 L
208 1841 L
214 1839 L
222 1839 L
228 1841 L
230 1843 L
CS M
231 1847 L
231 1855 L
230 1858 L
228 1860 L
222 1862 L
214 1862 M 210 1860 L
208 1858 L
206 1855 L
206 1847 L
208 1843 L
210 1841 L
214 1839 L
222 1839 M 226 1841 L
228 1843 L
230 1847 L
230 1855 L
228 1858 L
226 1860 L
222 1862 L
199 2240 M 210 2273 M 214 2275 L
220 2281 L
220 2240 L
218 2279 M 218 2240 L
210 2240 M 228 2240 L
2261 255 M 2261 2261 L
2261 255 M 2219 255 L
2261 355 M 2240 355 L
2261 455 M 2240 455 L
2261 556 M 2240 556 L
2261 656 M 2219 656 L
2261 756 M 2240 756 L
2261 857 M 2240 857 L
2261 957 M 2240 957 L
2261 1057 M 2219 1057 L
2261 1157 M 2240 1157 L
2261 1258 M 2240 1258 L
2261 1358 M 2240 1358 L
2261 1458 M 2219 1458 L
2261 1559 M 2240 1559 L
2261 1659 M 2240 1659 L
2261 1759 M 2240 1759 L
2261 1859 M 2219 1859 L
2261 1960 M 2240 1960 L
2261 2060 M 2240 2060 L
2261 2160 M 2240 2160 L
2261 2261 M 2219 2261 L
CS [6 12] 0 setdash M
346 280 M B
354 272 M 354 288 L
338 288 L
338 272 L
354 272 L
CF M
486 295 M B
494 287 M 494 303 L
478 303 L
478 287 L
494 287 L
CF M
626 315 M B
634 307 M 634 322 L
619 322 L
619 307 L
634 307 L
CF M
767 340 M B
775 332 M 775 348 L
759 348 L
759 332 L
775 332 L
CF M
907 372 M B
915 364 M 915 380 L
899 380 L
899 364 L
915 364 L
CF M
1047 411 M B
1055 403 M 1055 419 L
1039 419 L
1039 403 L
1055 403 L
CF M
1187 458 M B
1195 450 M 1195 466 L
1180 466 L
1180 450 L
1195 450 L
CF M
1328 513 M B
1336 505 M 1336 521 L
1320 521 L
1320 505 L
1336 505 L
CF M
1468 577 M B
1476 569 M 1476 585 L
1460 585 L
1460 569 L
1476 569 L
CF M
1608 651 M B
1616 643 M 1616 659 L
1600 659 L
1600 643 L
1616 643 L
CF M
1749 734 M B
1756 726 M 1756 742 L
1741 742 L
1741 726 L
1756 726 L
CF M
1889 829 M B
1897 821 M 1897 836 L
1881 836 L
1881 821 L
1897 821 L
CF M
2029 934 M B
2037 926 M 2037 942 L
2021 942 L
2021 926 L
2037 926 L
CF M
2169 1051 M B
2177 1043 M 2177 1059 L
2161 1059 L
2161 1043 L
2177 1043 L
CF M
346 527 M B
354 519 M 354 535 L
338 535 L
338 519 L
354 519 L
CF M
486 602 M B
494 595 M 494 610 L
478 610 L
478 595 L
494 595 L
CF M
626 690 M B
634 682 M 634 698 L
619 698 L
619 682 L
634 682 L
CF M
767 787 M B
775 779 M 775 795 L
759 795 L
759 780 L
775 780 L
CF M
907 896 M B
915 888 M 915 904 L
899 904 L
899 888 L
915 888 L
CF M
1047 1015 M B
1055 1007 M 1055 1023 L
1039 1023 L
1039 1007 L
1055 1007 L
CF M
1187 1131 M B
1195 1123 M 1195 1139 L
1180 1139 L
1180 1123 L
1195 1123 L
CF M
1328 1259 M B
1336 1251 M 1336 1266 L
1320 1266 L
1320 1251 L
1336 1251 L
CF M
1468 1392 M B
1476 1384 M 1476 1400 L
1460 1400 L
1460 1384 L
1476 1384 L
CF M
1608 1544 M B
1616 1536 M 1616 1552 L
1600 1552 L
1600 1536 L
1616 1536 L
CF M
1749 1647 M B
1756 1639 M 1756 1654 L
1741 1654 L
1741 1639 L
1756 1639 L
CF M
1889 1756 M B
1897 1748 M 1897 1764 L
1881 1764 L
1881 1748 L
1897 1748 L
CF M
2029 1854 M B
2037 1846 M 2037 1861 L
2021 1861 L
2021 1846 L
2037 1846 L
CF M
2169 1918 M B
2177 1910 M 2177 1926 L
2161 1926 L
2161 1910 L
2177 1910 L
CF M
346 312 M B
354 304 M 354 320 L
338 320 L
338 304 L
354 304 L
CF M
486 339 M B
494 332 M 494 347 L
478 347 L
478 332 L
494 332 L
CF M
626 374 M B
634 366 M 634 382 L
619 382 L
619 366 L
634 366 L
CF M
767 417 M B
775 409 M 775 425 L
759 425 L
759 409 L
775 409 L
CF M
907 469 M B
915 461 M 915 477 L
899 477 L
899 461 L
915 461 L
CF M
1047 531 M B
1055 523 M 1055 539 L
1039 539 L
1039 523 L
1055 523 L
CF M
1187 604 M B
1195 596 M 1195 612 L
1180 612 L
1180 596 L
1195 596 L
CF M
1328 688 M B
1336 680 M 1336 696 L
1320 696 L
1320 680 L
1336 680 L
CF M
1468 784 M B
1476 776 M 1476 792 L
1460 792 L
1460 776 L
1476 776 L
CF M
1608 894 M B
1616 886 M 1616 902 L
1600 902 L
1600 886 L
1616 886 L
CF M
1749 1018 M B
1756 1010 M 1756 1026 L
1741 1026 L
1741 1010 L
1756 1010 L
CF M
1889 1160 M B
1897 1152 M 1897 1168 L
1881 1168 L
1881 1152 L
1897 1152 L
CF M
2029 1322 M B
2037 1314 M 2037 1330 L
2021 1330 L
2021 1314 L
2037 1314 L
CF M
2169 1518 M B
2177 1510 M 2177 1526 L
2161 1526 L
2161 1510 L
2177 1510 L
CF M
346 362 M B
354 354 M 354 370 L
338 370 L
338 354 L
354 354 L
CF M
486 405 M B
494 397 M 494 413 L
478 413 L
478 397 L
494 397 L
CF M
626 459 M B
634 451 M 634 467 L
619 467 L
619 451 L
634 451 L
CF M
767 525 M B
775 517 M 775 533 L
759 533 L
759 517 L
775 517 L
CF M
907 603 M B
915 595 M 915 611 L
899 611 L
899 595 L
915 595 L
CF M
1047 696 M B
1055 688 M 1055 704 L
1039 704 L
1039 688 L
1055 688 L
CF M
1187 805 M B
1195 797 M 1195 813 L
1180 813 L
1180 797 L
1195 797 L
CF M
1328 934 M B
1336 926 M 1336 942 L
1320 942 L
1320 926 L
1336 926 L
CF M
1468 1089 M B
1476 1081 M 1476 1097 L
1460 1097 L
1460 1081 L
1476 1081 L
CF M
1608 1288 M B
1616 1280 M 1616 1296 L
1600 1296 L
1600 1280 L
1616 1280 L
CF M
CS [] 0 setdash M
346 527 M 486 602 L
626 690 L
767 787 L
907 896 L
1047 1015 L
1187 1131 L
1328 1259 L
1468 1392 L
1608 1544 L
1749 1647 L
1889 1756 L
2029 1854 L
2169 1918 L
CS [6 12] 0 setdash M
346 280 M 486 295 L
626 315 L
767 340 L
907 372 L
1047 411 L
1187 458 L
1328 513 L
1468 577 L
1608 651 L
1749 734 L
1889 829 L
2029 934 L
2169 1051 L
346 312 M 486 339 L
626 374 L
767 417 L
907 469 L
1047 531 L
1187 604 L
1328 688 L
1468 784 L
1608 894 L
1749 1018 L
1889 1160 L
2029 1322 L
2169 1518 L
346 362 M 486 405 L
626 459 L
767 525 L
907 603 L
1047 696 L
1187 805 L
1328 934 L
1468 1089 L
1608 1288 L
CS [] 0 setdash M
1105 115 M 1126 165 M 1126 115 L
1129 165 M 1129 115 L
1112 165 M 1110 151 L
1110 165 L
1146 165 L
1146 151 L
1143 165 L
1119 115 M 1136 115 L
1216 175 M 1211 170 L
1206 163 L
1201 153 L
1199 141 L
1199 132 L
1201 120 L
1206 110 L
1211 103 L
1216 98 L
1211 170 M 1206 161 L
1203 153 L
1201 141 L
1201 132 L
1203 120 L
1206 112 L
1211 103 L
1235 165 M 1235 115 L
1237 165 M 1252 122 L
1235 165 M 1252 115 L
1268 165 M 1252 115 L
1268 165 M 1268 115 L
1271 165 M 1271 115 L
1228 165 M 1237 165 L
1268 165 M 1278 165 L
1228 115 M 1242 115 L
1261 115 M 1278 115 L
1293 134 M 1321 134 L
1321 139 L
1319 144 L
1317 146 L
1312 149 L
1305 149 L
1297 146 L
1293 141 L
1290 134 L
1290 129 L
1293 122 L
1297 117 L
1305 115 L
1309 115 L
1317 117 L
1321 122 L
1319 134 M 1319 141 L
1317 146 L
1305 149 M 1300 146 L
1295 141 L
1293 134 L
1293 129 L
1295 122 L
1300 117 L
1305 115 L
1336 165 M 1353 115 L
1338 165 M 1353 122 L
1370 165 M 1353 115 L
1331 165 M 1345 165 L
1360 165 M 1374 165 L
1384 175 M 1389 170 L
1394 163 L
1398 153 L
1401 141 L
1401 132 L
1398 120 L
1394 110 L
1389 103 L
1384 98 L
1389 170 M 1394 161 L
1396 153 L
1398 141 L
1398 132 L
1396 120 L
1394 112 L
1389 103 L
CS [6 12] 0 setdash M
CS [] 0 setdash M
102 1090 M 69 1102 M 102 1102 L
69 1104 M 102 1104 L
76 1104 M 71 1109 L
69 1116 L
69 1121 L
71 1128 L
76 1131 L
102 1131 L
69 1121 M 71 1126 L
76 1128 L
102 1128 L
69 1095 M 69 1104 L
102 1095 M 102 1111 L
102 1121 M 102 1138 L
42 1208 M 47 1203 L
54 1198 L
64 1193 L
76 1191 L
85 1191 L
97 1193 L
107 1198 L
114 1203 L
119 1208 L
47 1203 M 57 1198 L
64 1196 L
76 1193 L
85 1193 L
97 1196 L
105 1198 L
114 1203 L
54 1239 M 57 1237 L
59 1239 L
57 1241 L
54 1241 L
52 1239 L
52 1234 L
54 1229 L
59 1227 L
102 1227 L
52 1234 M 54 1232 L
59 1229 L
102 1229 L
69 1220 M 69 1239 L
102 1220 M 102 1237 L
69 1258 M 102 1258 L
69 1261 M 102 1261 L
76 1261 M 71 1265 L
69 1273 L
69 1277 L
71 1285 L
76 1287 L
102 1287 L
69 1277 M 71 1282 L
76 1285 L
102 1285 L
76 1287 M 71 1292 L
69 1299 L
69 1304 L
71 1311 L
76 1314 L
102 1314 L
69 1304 M 71 1309 L
76 1311 L
102 1311 L
69 1251 M 69 1261 L
102 1251 M 102 1268 L
102 1277 M 102 1294 L
102 1304 M 102 1321 L
66 1331 M 66 1357 L
55 1369 M 56 1370 L
57 1369 L
56 1367 L
55 1367 L
52 1369 L
50 1370 L
49 1375 L
49 1380 L
50 1385 L
53 1386 L
57 1386 L
60 1385 L
62 1380 L
62 1376 L
49 1380 M 50 1383 L
53 1385 L
57 1385 L
60 1383 L
62 1380 L
63 1383 L
66 1386 L
69 1388 L
73 1388 L
76 1386 L
78 1385 L
79 1380 L
79 1375 L
78 1370 L
76 1369 L
73 1367 L
72 1367 L
70 1369 L
72 1370 L
73 1369 L
65 1385 M 69 1386 L
73 1386 L
76 1385 L
78 1383 L
79 1380 L
42 1399 M 47 1404 L
54 1409 L
64 1414 L
76 1416 L
85 1416 L
97 1414 L
107 1409 L
114 1404 L
119 1399 L
47 1404 M 57 1409 L
64 1411 L
76 1414 L
85 1414 L
97 1411 L
105 1409 L
114 1404 L
CS [6 12] 0 setdash M
1749 1819 M CS [] 0 setdash M
1648 1819 M 1665 1853 M 1651 1802 L
1667 1853 M 1653 1802 L
1665 1846 M 1663 1831 L
1663 1824 L
1667 1819 L
1672 1819 L
1677 1822 L
1682 1826 L
1687 1834 L
1691 1853 M 1684 1826 L
1684 1822 L
1687 1819 L
1694 1819 L
1699 1824 L
1701 1829 L
1694 1853 M 1687 1826 L
1687 1822 L
1689 1819 L
1713 1848 M 1756 1848 L
1713 1834 M 1756 1834 L
1783 1853 M 1769 1802 L
1785 1853 M 1771 1802 L
1783 1846 M 1781 1831 L
1781 1824 L
1785 1819 L
1790 1819 L
1795 1822 L
1800 1826 L
1805 1834 L
1809 1853 M 1802 1826 L
1802 1822 L
1805 1819 L
1812 1819 L
1817 1824 L
1819 1829 L
1812 1853 M 1805 1826 L
1805 1822 L
1807 1819 L
1843 1812 M 1842 1811 L
1843 1809 L
1845 1811 L
1845 1812 L
1842 1815 L
1839 1816 L
1834 1816 L
1830 1815 L
1827 1812 L
1826 1808 L
CS M
1826 1805 L
1827 1800 L
1830 1798 L
1834 1796 L
1837 1796 L
1842 1798 L
1845 1800 L
1834 1816 M 1832 1815 L
1829 1812 L
1827 1808 L
1827 1805 L
1829 1800 L
1832 1798 L
1834 1796 L
CS [6 12] 0 setdash M
1749 816 M CS [] 0 setdash M
1665 816 M 1682 850 M 1668 800 L
1685 850 M 1670 800 L
1682 843 M 1680 828 L
1680 821 L
1685 816 L
1690 816 L
1694 819 L
1699 824 L
1704 831 L
1709 850 M 1702 824 L
1702 819 L
1704 816 L
1711 816 L
1716 821 L
1718 826 L
1711 850 M 1704 824 L
1704 819 L
1706 816 L
1730 845 M 1774 845 L
1730 831 M 1774 831 L
1805 867 M 1798 865 L
1793 857 L
1791 845 L
1791 838 L
1793 826 L
1798 819 L
1805 816 L
1810 816 L
1817 819 L
1822 826 L
1824 838 L
1824 845 L
1822 857 L
1817 865 L
1810 867 L
1805 867 L
1800 865 L
1798 862 L
1795 857 L
1793 845 L
1793 838 L
1795 826 L
1798 821 L
1800 819 L
1805 816 L
1810 816 M 1815 819 L
1817 821 L
1819 826 L
1822 838 L
1822 845 L
1819 857 L
1817 862 L
1815 865 L
1810 867 L
CS [6 12] 0 setdash M
1749 1097 M CS [] 0 setdash M
1641 1097 M 1658 1131 M 1644 1080 L
1661 1131 M 1646 1080 L
1658 1124 M 1656 1109 L
1656 1102 L
1661 1097 L
1665 1097 L
1670 1100 L
1675 1104 L
1680 1112 L
1685 1131 M 1677 1104 L
1677 1100 L
1680 1097 L
1687 1097 L
1692 1102 L
1694 1107 L
1687 1131 M 1680 1104 L
1680 1100 L
1682 1097 L
1706 1126 M 1750 1126 L
1706 1112 M 1750 1112 L
1795 1141 M 1793 1138 L
1795 1136 L
1798 1138 L
1798 1141 L
1795 1145 L
1791 1148 L
1783 1148 L
1776 1145 L
1771 1141 L
1769 1136 L
1767 1126 L
1767 1112 L
1769 1104 L
1774 1100 L
1781 1097 L
1786 1097 L
1793 1100 L
1798 1104 L
1800 1112 L
1800 1114 L
1798 1121 L
1793 1126 L
1786 1128 L
1783 1128 L
1776 1126 L
1771 1121 L
1769 1114 L
1783 1148 M 1779 1145 L
1774 1141 L
1771 1136 L
1769 1126 L
1769 1112 L
1771 1104 L
1776 1100 L
1781 1097 L
1786 1097 M 1791 1100 L
1795 1104 L
1798 1112 L
1798 1114 L
1795 1121 L
1791 1126 L
1786 1128 L
1829 1148 M 1822 1145 L
1817 1138 L
1815 1126 L
1815 1119 L
1817 1107 L
1822 1100 L
1829 1097 L
1834 1097 L
1841 1100 L
1846 1107 L
1848 1119 L
1848 1126 L
1846 1138 L
1841 1145 L
1834 1148 L
1829 1148 L
1824 1145 L
1822 1143 L
1819 1138 L
1817 1126 L
1817 1119 L
1819 1107 L
1822 1102 L
1824 1100 L
1829 1097 L
1834 1097 M 1839 1100 L
1841 1102 L
1844 1107 L
1846 1119 L
1846 1126 L
1844 1138 L
1841 1143 L
1839 1145 L
1834 1148 L
CS [6 12] 0 setdash M
1749 1338 M CS [] 0 setdash M
1617 1338 M 1634 1372 M 1620 1321 L
1637 1372 M 1622 1321 L
1634 1364 M 1632 1350 L
1632 1343 L
1637 1338 L
1641 1338 L
1646 1340 L
1651 1345 L
1656 1352 L
1661 1372 M 1653 1345 L
1653 1340 L
1656 1338 L
1663 1338 L
1668 1343 L
1670 1348 L
1663 1372 M 1656 1345 L
1656 1340 L
1658 1338 L
1682 1367 M 1726 1367 L
1682 1352 M 1726 1352 L
1750 1379 M 1755 1381 L
1762 1388 L
1762 1338 L
1759 1386 M 1759 1338 L
1750 1338 M 1771 1338 L
1805 1388 M 1798 1386 L
1793 1379 L
1791 1367 L
1791 1360 L
1793 1348 L
1798 1340 L
1805 1338 L
1810 1338 L
1817 1340 L
1822 1348 L
1824 1360 L
1824 1367 L
1822 1379 L
1817 1386 L
1810 1388 L
1805 1388 L
1800 1386 L
1798 1384 L
1795 1379 L
1793 1367 L
1793 1360 L
1795 1348 L
1798 1343 L
1800 1340 L
1805 1338 L
1810 1338 M 1815 1340 L
1817 1343 L
1819 1348 L
1822 1360 L
1822 1367 L
1819 1379 L
1817 1384 L
1815 1386 L
1810 1388 L
1853 1388 M 1846 1386 L
1841 1379 L
1839 1367 L
1839 1360 L
1841 1348 L
1846 1340 L
1853 1338 L
1858 1338 L
1865 1340 L
1870 1348 L
1872 1360 L
1872 1367 L
1870 1379 L
1865 1386 L
1858 1388 L
1853 1388 L
1848 1386 L
1846 1384 L
1844 1379 L
1841 1367 L
1841 1360 L
1844 1348 L
1846 1343 L
1848 1340 L
1853 1338 L
1858 1338 M 1863 1340 L
1865 1343 L
1868 1348 L
1870 1360 L
1870 1367 L
1868 1379 L
1865 1384 L
1863 1386 L
1858 1388 L
CS [6 12] 0 setdash M
stroke
grestore
showpage
end

 20 dict begin
72 300 div dup scale
1 setlinejoin 0 setlinecap
/Helvetica findfont 55 scalefont setfont
/B { stroke newpath } def /F { moveto 0 setlinecap} def
/C { CS M 1 1 3 { pop 3 1 roll 255 div } for SET_COLOUR } def
/CS { currentpoint stroke } def
/CF { currentpoint fill } def
/L { lineto } def /M { moveto } def
/P { moveto 0 1 rlineto stroke } def
/T { 1 setlinecap show } def
errordict /nocurrentpoint { pop 0 0 M currentpoint } put
/SET_COLOUR { pop pop pop } def
 80 600 translate
gsave
CS [6 12] 0 setdash M
CS M 2 setlinewidth
/P { moveto 0 2.05 rlineto stroke } def
 0 0 0 C
CS [] 0 setdash M
255 255 M 2261 255 L
282 255 M 282 280 L
421 255 M 421 280 L
559 255 M 559 306 L
697 255 M 697 280 L
836 255 M 836 280 L
974 255 M 974 280 L
1112 255 M 1112 280 L
1251 255 M 1251 306 L
1389 255 M 1389 280 L
1527 255 M 1527 280 L
1666 255 M 1666 280 L
1804 255 M 1804 280 L
1942 255 M 1942 306 L
2081 255 M 2081 280 L
2219 255 M 2219 280 L
487 182 M 501 223 M 506 226 L
513 233 L
513 182 L
511 230 M 511 182 L
501 182 M 523 182 L
557 233 M 549 230 L
545 223 L
542 211 L
542 204 L
545 192 L
549 185 L
557 182 L
561 182 L
569 185 L
574 192 L
576 204 L
576 211 L
574 223 L
569 230 L
561 233 L
557 233 L
552 230 L
549 228 L
547 223 L
545 211 L
545 204 L
547 192 L
549 187 L
552 185 L
557 182 L
561 182 M 566 185 L
569 187 L
571 192 L
574 204 L
574 211 L
571 223 L
569 228 L
566 230 L
561 233 L
605 233 M 598 230 L
593 223 L
590 211 L
590 204 L
593 192 L
598 185 L
605 182 L
610 182 L
617 185 L
622 192 L
624 204 L
624 211 L
622 223 L
617 230 L
610 233 L
605 233 L
600 230 L
598 228 L
595 223 L
593 211 L
593 204 L
595 192 L
598 187 L
600 185 L
605 182 L
610 182 M 614 185 L
617 187 L
619 192 L
622 204 L
622 211 L
619 223 L
617 228 L
614 230 L
610 233 L
1179 182 M 1193 223 M 1198 226 L
1205 233 L
1205 182 L
1203 230 M 1203 182 L
1193 182 M 1215 182 L
1239 233 M 1234 209 L
1239 214 L
1246 216 L
1253 216 L
1260 214 L
1265 209 L
1268 202 L
1268 197 L
1265 190 L
1260 185 L
1253 182 L
1246 182 L
1239 185 L
1236 187 L
1234 192 L
1234 194 L
1236 197 L
1239 194 L
1236 192 L
1253 216 M 1258 214 L
1263 209 L
1265 202 L
1265 197 L
1263 190 L
1258 185 L
1253 182 L
1239 233 M 1263 233 L
1239 230 M 1251 230 L
1263 233 L
1296 233 M 1289 230 L
1284 223 L
1282 211 L
1282 204 L
1284 192 L
1289 185 L
1296 182 L
1301 182 L
1309 185 L
1313 192 L
1316 204 L
1316 211 L
1313 223 L
1309 230 L
1301 233 L
1296 233 L
1292 230 L
1289 228 L
1287 223 L
1284 211 L
1284 204 L
1287 192 L
1289 187 L
1292 185 L
1296 182 L
1301 182 M 1306 185 L
1309 187 L
1311 192 L
1313 204 L
1313 211 L
1311 223 L
1309 228 L
1306 230 L
1301 233 L
1870 182 M 1880 223 M 1882 221 L
1880 218 L
1877 221 L
1877 223 L
1880 228 L
1882 230 L
1889 233 L
1899 233 L
1906 230 L
1909 228 L
1911 223 L
1911 218 L
1909 214 L
1901 209 L
1889 204 L
1885 202 L
1880 197 L
1877 190 L
1877 182 L
1899 233 M 1904 230 L
1906 228 L
1909 223 L
1909 218 L
1906 214 L
1899 209 L
1889 204 L
1877 187 M 1880 190 L
1885 190 L
1897 185 L
1904 185 L
1909 187 L
1911 190 L
1885 190 M 1897 182 L
1906 182 L
1909 185 L
1911 190 L
1911 194 L
1940 233 M 1933 230 L
1928 223 L
1925 211 L
1925 204 L
1928 192 L
1933 185 L
1940 182 L
1945 182 L
1952 185 L
1957 192 L
1959 204 L
1959 211 L
1957 223 L
1952 230 L
1945 233 L
1940 233 L
1935 230 L
1933 228 L
1930 223 L
1928 211 L
1928 204 L
1930 192 L
1933 187 L
1935 185 L
1940 182 L
1945 182 M 1950 185 L
1952 187 L
1954 192 L
1957 204 L
1957 211 L
1954 223 L
1952 228 L
1950 230 L
1945 233 L
1988 233 M 1981 230 L
1976 223 L
1974 211 L
1974 204 L
1976 192 L
1981 185 L
1988 182 L
1993 182 L
2000 185 L
2005 192 L
2007 204 L
2007 211 L
2005 223 L
2000 230 L
1993 233 L
1988 233 L
1983 230 L
1981 228 L
1978 223 L
1976 211 L
1976 204 L
1978 192 L
1981 187 L
1983 185 L
1988 182 L
1993 182 M 1998 185 L
2000 187 L
2002 192 L
2005 204 L
2005 211 L
2002 223 L
2000 228 L
1998 230 L
1993 233 L
255 2261 M 2261 2261 L
282 2261 M 282 2235 L
421 2261 M 421 2235 L
559 2261 M 559 2209 L
697 2261 M 697 2235 L
836 2261 M 836 2235 L
974 2261 M 974 2235 L
1112 2261 M 1112 2235 L
1251 2261 M 1251 2209 L
1389 2261 M 1389 2235 L
1527 2261 M 1527 2235 L
1666 2261 M 1666 2235 L
1804 2261 M 1804 2235 L
1942 2261 M 1942 2209 L
2081 2261 M 2081 2235 L
2219 2261 M 2219 2235 L
255 255 M 255 2261 L
255 255 M 306 255 L
255 335 M 280 335 L
255 415 M 280 415 L
255 495 M 280 495 L
255 576 M 280 576 L
255 656 M 306 656 L
255 736 M 280 736 L
255 816 M 280 816 L
255 897 M 280 897 L
255 977 M 280 977 L
255 1057 M 306 1057 L
255 1137 M 280 1137 L
255 1218 M 280 1218 L
255 1298 M 280 1298 L
255 1378 M 280 1378 L
255 1458 M 306 1458 L
255 1538 M 280 1538 L
255 1619 M 280 1619 L
255 1699 M 280 1699 L
255 1779 M 280 1779 L
255 1859 M 306 1859 L
CS M
255 1940 M 280 1940 L
255 2020 M 280 2020 L
255 2100 M 280 2100 L
255 2180 M 280 2180 L
255 2261 M 306 2261 L
185 229 M 206 280 M 199 278 L
194 270 L
192 258 L
192 251 L
194 239 L
199 232 L
206 229 L
211 229 L
218 232 L
223 239 L
226 251 L
226 258 L
223 270 L
218 278 L
211 280 L
206 280 L
202 278 L
199 275 L
197 270 L
194 258 L
194 251 L
197 239 L
199 234 L
202 232 L
206 229 L
211 229 M 216 232 L
218 234 L
221 239 L
223 251 L
223 258 L
221 270 L
218 275 L
216 278 L
211 280 L
185 631 M 199 672 M 204 674 L
211 681 L
211 631 L
209 679 M 209 631 L
199 631 M 221 631 L
185 1032 M 194 1073 M 197 1070 L
194 1068 L
192 1070 L
192 1073 L
194 1078 L
197 1080 L
204 1082 L
214 1082 L
221 1080 L
223 1078 L
226 1073 L
226 1068 L
223 1063 L
216 1058 L
204 1053 L
199 1051 L
194 1046 L
192 1039 L
192 1032 L
214 1082 M 218 1080 L
221 1078 L
223 1073 L
223 1068 L
221 1063 L
214 1058 L
204 1053 L
192 1037 M 194 1039 L
199 1039 L
211 1034 L
218 1034 L
223 1037 L
226 1039 L
199 1039 M 211 1032 L
221 1032 L
223 1034 L
226 1039 L
226 1044 L
185 1433 M 194 1474 M 197 1471 L
194 1469 L
192 1471 L
192 1474 L
194 1479 L
197 1481 L
204 1483 L
214 1483 L
221 1481 L
223 1476 L
223 1469 L
221 1464 L
214 1462 L
206 1462 L
214 1483 M 218 1481 L
221 1476 L
221 1469 L
218 1464 L
214 1462 L
218 1459 L
223 1455 L
226 1450 L
226 1443 L
223 1438 L
221 1435 L
214 1433 L
204 1433 L
197 1435 L
194 1438 L
192 1443 L
192 1445 L
194 1447 L
197 1445 L
194 1443 L
221 1457 M 223 1450 L
223 1443 L
221 1438 L
218 1435 L
214 1433 L
185 1834 M 214 1880 M 214 1834 L
216 1885 M 216 1834 L
216 1885 M 190 1849 L
228 1849 L
206 1834 M 223 1834 L
185 2235 M 197 2286 M 192 2262 L
197 2267 L
204 2269 L
211 2269 L
218 2267 L
223 2262 L
226 2254 L
226 2250 L
223 2242 L
218 2238 L
211 2235 L
204 2235 L
197 2238 L
194 2240 L
192 2245 L
192 2247 L
194 2250 L
197 2247 L
194 2245 L
211 2269 M 216 2267 L
221 2262 L
223 2254 L
223 2250 L
221 2242 L
216 2238 L
211 2235 L
197 2286 M 221 2286 L
197 2283 M 209 2283 L
221 2286 L
2261 255 M 2261 2261 L
2261 255 M 2209 255 L
2261 335 M 2235 335 L
2261 415 M 2235 415 L
2261 495 M 2235 495 L
2261 576 M 2235 576 L
2261 656 M 2209 656 L
2261 736 M 2235 736 L
2261 816 M 2235 816 L
2261 897 M 2235 897 L
2261 977 M 2235 977 L
2261 1057 M 2209 1057 L
2261 1137 M 2235 1137 L
2261 1218 M 2235 1218 L
2261 1298 M 2235 1298 L
2261 1378 M 2235 1378 L
2261 1458 M 2209 1458 L
2261 1538 M 2235 1538 L
2261 1619 M 2235 1619 L
2261 1699 M 2235 1699 L
2261 1779 M 2235 1779 L
2261 1859 M 2209 1859 L
2261 1940 M 2235 1940 L
2261 2020 M 2235 2020 L
2261 2100 M 2235 2100 L
2261 2180 M 2235 2180 L
2261 2261 M 2209 2261 L
CS [6 12] 0 setdash M
CS [] 0 setdash M
282 2187 M B
292 2177 M 292 2196 L
273 2196 L
273 2177 L
292 2177 L
CF M
421 2116 M B
431 2107 M 431 2126 L
411 2126 L
411 2107 L
431 2107 L
CF M
559 2059 M B
569 2049 M 569 2069 L
549 2069 L
549 2049 L
569 2049 L
CF M
697 2011 M B
707 2001 M 707 2021 L
687 2021 L
687 2001 L
707 2001 L
CF M
836 1969 M B
846 1959 M 846 1979 L
826 1979 L
826 1959 L
846 1959 L
CF M
974 1932 M B
984 1923 M 984 1942 L
964 1942 L
964 1923 L
984 1923 L
CF M
1112 1900 M B
1122 1890 M 1122 1910 L
1102 1910 L
1102 1890 L
1122 1890 L
CF M
1251 1871 M B
1261 1861 M 1261 1881 L
1241 1881 L
1241 1861 L
1261 1861 L
CF M
1389 1846 M B
1399 1836 M 1399 1856 L
1379 1856 L
1379 1836 L
1399 1836 L
CF M
1527 1822 M B
1537 1812 M 1537 1832 L
1517 1832 L
1517 1812 L
1537 1812 L
CF M
1666 1801 M B
1676 1791 M 1676 1811 L
1656 1811 L
1656 1791 L
1676 1791 L
CF M
1804 1782 M B
1814 1772 M 1814 1792 L
1794 1792 L
1794 1772 L
1814 1772 L
CF M
1942 1765 M B
1952 1755 M 1952 1774 L
1932 1774 L
1932 1755 L
1952 1755 L
CF M
2081 1748 M B
2091 1738 M 2091 1758 L
2071 1758 L
2071 1738 L
2091 1738 L
CF M
2219 1732 M B
2229 1722 M 2229 1742 L
2209 1742 L
2209 1722 L
2229 1722 L
CF M
282 2187 M 421 2116 L
559 2059 L
697 2011 L
836 1969 L
974 1932 L
1112 1900 L
1251 1871 L
1389 1846 L
1527 1822 L
1666 1801 L
1804 1782 L
1942 1765 L
2081 1748 L
2219 1732 L
282 2075 M B
292 2065 M 292 2085 L
273 2085 L
273 2065 L
292 2065 L
CF M
421 2015 M B
431 2005 M 431 2025 L
411 2025 L
411 2005 L
431 2005 L
CF M
559 1964 M B
569 1955 M 569 1974 L
549 1974 L
549 1955 L
569 1955 L
CF M
697 1922 M B
707 1912 M 707 1932 L
687 1932 L
687 1912 L
707 1912 L
CF M
836 1884 M B
846 1874 M 846 1894 L
826 1894 L
826 1874 L
846 1874 L
CF M
974 1851 M B
984 1841 M 984 1861 L
964 1861 L
964 1841 L
984 1841 L
CF M
1112 1822 M B
1122 1812 M 1122 1832 L
1102 1832 L
1102 1812 L
1122 1812 L
CF M
1251 1795 M B
1261 1785 M 1261 1805 L
1241 1805 L
1241 1785 L
1261 1785 L
CF M
1389 1771 M B
1399 1761 M 1399 1781 L
1379 1781 L
1379 1761 L
1399 1761 L
CF M
1527 1749 M B
1537 1739 M 1537 1759 L
1517 1759 L
1517 1739 L
1537 1739 L
CF M
1666 1729 M B
1676 1719 M 1676 1739 L
1656 1739 L
1656 1719 L
1676 1719 L
CF M
1804 1710 M B
1814 1700 M 1814 1720 L
1794 1720 L
1794 1700 L
1814 1700 L
CF M
1942 1692 M B
1952 1682 M 1952 1702 L
1932 1702 L
1932 1682 L
1952 1682 L
CF M
2081 1675 M B
2091 1665 M 2091 1685 L
2071 1685 L
2071 1665 L
2091 1665 L
CF M
2219 1656 M B
2229 1647 M 2229 1666 L
2209 1666 L
2209 1647 L
2229 1647 L
CF M
282 2075 M 421 2015 L
559 1964 L
697 1922 L
836 1884 L
974 1851 L
1112 1822 L
1251 1795 L
1389 1771 L
1527 1749 L
1666 1729 L
1804 1710 L
1942 1692 L
2081 1675 L
2219 1656 L
282 1960 M B
292 1950 M 292 1970 L
273 1970 L
273 1950 L
292 1950 L
CF M
421 1909 M B
431 1899 M 431 1919 L
411 1919 L
411 1899 L
431 1899 L
CF M
559 1865 M B
569 1855 M 569 1875 L
549 1875 L
549 1855 L
569 1855 L
CF M
697 1828 M B
707 1818 M 707 1838 L
687 1838 L
687 1818 L
707 1818 L
CF M
836 1795 M B
846 1785 M 846 1805 L
826 1805 L
826 1785 L
846 1785 L
CF M
974 1765 M B
984 1755 M 984 1775 L
964 1775 L
964 1755 L
984 1755 L
CF M
1112 1738 M B
1122 1728 M 1122 1747 L
1102 1747 L
1102 1728 L
1122 1728 L
CF M
1251 1713 M B
1261 1703 M 1261 1723 L
1241 1723 L
1241 1703 L
1261 1703 L
CF M
1389 1690 M B
1399 1680 M 1399 1700 L
1379 1700 L
1379 1680 L
1399 1680 L
CF M
1527 1668 M B
1537 1658 M 1537 1678 L
1517 1678 L
1517 1658 L
1537 1658 L
CF M
1666 1648 M B
1676 1638 M 1676 1658 L
1656 1658 L
1656 1638 L
1676 1638 L
CF M
1804 1628 M B
1814 1618 M 1814 1638 L
1794 1638 L
1794 1618 L
1814 1618 L
CF M
1942 1608 M B
1952 1598 M 1952 1618 L
1932 1618 L
1932 1598 L
1952 1598 L
CF M
2081 1586 M B
2091 1576 M 2091 1596 L
2071 1596 L
2071 1576 L
2091 1576 L
CF M
2219 1560 M B
2229 1550 M 2229 1570 L
2209 1570 L
2209 1550 L
2229 1550 L
CF M
282 1960 M 421 1909 L
559 1865 L
697 1828 L
836 1795 L
974 1765 L
1112 1738 L
1251 1713 L
1389 1690 L
1527 1668 L
1666 1648 L
1804 1628 L
1942 1608 L
2081 1586 L
2219 1560 L
282 1839 M B
292 1829 M 292 1849 L
273 1849 L
273 1829 L
292 1829 L
CF M
421 1797 M B
431 1787 M 431 1807 L
411 1807 L
411 1787 L
431 1787 L
CF M
559 1760 M B
569 1750 M 569 1770 L
549 1770 L
549 1750 L
569 1750 L
CF M
697 1728 M B
707 1718 M 707 1737 L
687 1737 L
687 1718 L
707 1718 L
CF M
836 1698 M B
846 1688 M 846 1708 L
826 1708 L
826 1688 L
846 1688 L
CF M
974 1671 M B
984 1661 M 984 1681 L
964 1681 L
964 1661 L
984 1661 L
CF M
1112 1645 M B
1122 1635 M 1122 1655 L
1102 1655 L
1102 1635 L
1122 1635 L
CF M
1251 1621 M B
1261 1611 M 1261 1631 L
1241 1631 L
1241 1611 L
1261 1611 L
CF M
1389 1598 M B
1399 1588 M 1399 1608 L
1379 1608 L
1379 1588 L
1399 1588 L
CF M
1527 1576 M B
1537 1566 M 1537 1586 L
1517 1586 L
1517 1566 L
1537 1566 L
CF M
1666 1553 M B
1676 1543 M 1676 1563 L
1656 1563 L
1656 1543 L
1676 1543 L
CF M
1804 1529 M B
1814 1519 M 1814 1539 L
1794 1539 L
1794 1519 L
1814 1519 L
CF M
1942 1501 M B
1952 1491 M 1952 1511 L
1932 1511 L
1932 1491 L
1952 1491 L
CF M
2081 1462 M B
2091 1452 M 2091 1472 L
2071 1472 L
2071 1452 L
2091 1452 L
CF M
2219 1338 M B
2229 1328 M 2229 1348 L
2209 1348 L
2209 1328 L
2229 1328 L
CF M
282 1839 M 421 1797 L
559 1760 L
697 1728 L
836 1698 L
974 1671 L
1112 1645 L
1251 1621 L
1389 1598 L
1527 1576 L
1666 1553 L
1804 1529 L
1942 1501 L
2081 1462 L
2219 1338 L
282 1710 M B
292 1700 M 292 1719 L
273 1719 L
273 1700 L
292 1700 L
CF M
421 1676 M B
431 1666 M 431 1685 L
411 1685 L
411 1666 L
431 1666 L
CF M
559 1645 M B
569 1635 M 569 1655 L
549 1655 L
549 1635 L
569 1635 L
CF M
697 1617 M B
707 1607 M 707 1627 L
687 1627 L
687 1607 L
707 1607 L
CF M
836 1590 M B
846 1580 M 846 1600 L
826 1600 L
826 1580 L
846 1580 L
CF M
974 1565 M B
984 1555 M 984 1574 L
964 1574 L
964 1555 L
984 1555 L
CF M
1112 1539 M B
1122 1530 M 1122 1549 L
1102 1549 L
1102 1530 L
1122 1530 L
CF M
1251 1515 M B
1261 1505 M 1261 1525 L
1241 1525 L
1241 1505 L
1261 1505 L
CF M
1389 1489 M B
1399 1479 M 1399 1499 L
1379 1499 L
1379 1479 L
1399 1479 L
CF M
1527 1462 M B
1537 1452 M 1537 1472 L
1517 1472 L
1517 1452 L
1537 1452 L
CF M
1666 1430 M B
1676 1420 M 1676 1440 L
1656 1440 L
1656 1420 L
1676 1420 L
CF M
1804 1387 M B
1814 1377 M 1814 1397 L
1794 1397 L
1794 1377 L
1814 1377 L
CF M
1942 1298 M B
1952 1288 M 1952 1307 L
1932 1307 L
1932 1288 L
1952 1288 L
CF M
282 1710 M 421 1676 L
559 1645 L
697 1617 L
836 1590 L
974 1565 L
1112 1539 L
1251 1515 L
1389 1489 L
1527 1462 L
1666 1430 L
1804 1387 L
1942 1298 L
282 1564 M B
292 1554 M 292 1574 L
273 1574 L
273 1554 L
292 1554 L
CF M
421 1538 M B
431 1528 M 431 1548 L
411 1548 L
411 1528 L
431 1528 L
CF M
559 1513 M B
569 1503 M 569 1523 L
549 1523 L
549 1503 L
569 1503 L
CF M
697 1488 M B
707 1478 M 707 1498 L
687 1498 L
687 1478 L
707 1478 L
CF M
836 1463 M B
846 1453 M 846 1473 L
826 1473 L
826 1453 L
846 1453 L
CF M
974 1436 M B
984 1426 M 984 1446 L
964 1446 L
964 1426 L
984 1426 L
CF M
1112 1408 M B
1122 1398 M 1122 1418 L
1102 1418 L
1102 1398 L
1122 1398 L
CF M
1251 1376 M B
1261 1366 M 1261 1386 L
1241 1386 L
1241 1366 L
1261 1366 L
CF M
1389 1337 M B
1399 1327 M 1399 1347 L
1379 1347 L
1379 1327 L
1399 1327 L
CF M
1527 1281 M B
1537 1271 M 1537 1291 L
1517 1291 L
1517 1271 L
1537 1271 L
CF M
282 1564 M 421 1538 L
559 1513 L
697 1488 L
836 1463 L
974 1436 L
1112 1408 L
1251 1376 L
1389 1337 L
1527 1281 L
282 1383 M B
292 1373 M 292 1393 L
273 1393 L
273 1373 L
292 1373 L
CF M
421 1364 M B
431 1354 M 431 1374 L
411 1374 L
411 1354 L
431 1354 L
CF M
559 1342 M B
569 1332 M 569 1352 L
549 1352 L
549 1332 L
569 1332 L
CF M
697 1315 M B
707 1305 M 707 1325 L
687 1325 L
687 1305 L
707 1305 L
CF M
836 1283 M B
846 1273 M 846 1293 L
826 1293 L
826 1273 L
846 1273 L
CF M
974 1239 M B
984 1229 M 984 1249 L
964 1249 L
964 1229 L
984 1229 L
CF M
1112 1160 M B
1122 1150 M 1122 1170 L
1102 1170 L
1102 1150 L
1122 1150 L
CF M
282 1383 M 421 1364 L
559 1342 L
697 1315 L
836 1283 L
974 1239 L
1112 1160 L
282 1639 M B
292 1629 M 292 1649 L
273 1649 L
273 1629 L
292 1629 L
CF M
421 1610 M B
431 1600 M 431 1619 L
411 1619 L
411 1600 L
431 1600 L
CF M
559 1582 M B
569 1572 M 569 1592 L
549 1592 L
549 1572 L
569 1572 L
CF M
697 1556 M B
707 1546 M 707 1566 L
687 1566 L
687 1546 L
707 1546 L
CF M
836 1530 M B
846 1520 M 846 1540 L
826 1540 L
826 1520 L
846 1520 L
CF M
974 1504 M B
984 1494 M 984 1514 L
964 1514 L
964 1495 L
984 1495 L
CF M
1112 1478 M B
1122 1469 M 1122 1488 L
1102 1488 L
1102 1469 L
1122 1469 L
CF M
1251 1451 M B
1261 1442 M 1261 1461 L
1241 1461 L
1241 1442 L
1261 1442 L
CF M
1389 1422 M B
1399 1412 M 1399 1432 L
1379 1432 L
1379 1412 L
1399 1412 L
CF M
1527 1387 M B
1537 1377 M 1537 1397 L
1517 1397 L
1517 1378 L
1537 1378 L
CF M
1666 1340 M B
1676 1330 M 1676 1350 L
1656 1350 L
1656 1330 L
1676 1330 L
CF M
1804 1231 M B
1814 1221 M 1814 1241 L
1794 1241 L
1794 1221 L
1814 1221 L
CF M
282 1639 M 421 1610 L
559 1582 L
697 1556 L
836 1530 L
974 1504 L
1112 1478 L
1251 1451 L
1389 1422 L
1527 1387 L
1666 1340 L
1804 1231 L
282 1480 M B
292 1470 M 292 1490 L
273 1490 L
273 1470 L
292 1470 L
CF M
421 1458 M B
431 1448 M 431 1468 L
411 1468 L
411 1448 L
431 1448 L
CF M
559 1436 M B
569 1426 M 569 1445 L
549 1445 L
549 1426 L
569 1426 L
CF M
697 1411 M B
707 1401 M 707 1421 L
687 1421 L
687 1401 L
707 1401 L
CF M
836 1385 M B
846 1375 M 846 1394 L
826 1394 L
826 1375 L
846 1375 L
CF M
974 1355 M B
984 1345 M 984 1365 L
964 1365 L
964 1345 L
984 1345 L
CF M
1112 1319 M B
1122 1309 M 1122 1329 L
1102 1329 L
1102 1309 L
1122 1309 L
CF M
1251 1273 M B
1261 1263 M 1261 1283 L
1241 1283 L
1241 1263 L
1261 1263 L
CF M
1389 1188 M B
1399 1178 M 1399 1198 L
1379 1198 L
1379 1178 L
1399 1178 L
CF M
282 1480 M 421 1458 L
559 1436 L
697 1411 L
836 1385 L
974 1355 L
1112 1319 L
1251 1273 L
1389 1188 L
282 1253 M B
292 1243 M 292 1263 L
273 1263 L
273 1243 L
292 1243 L
CF M
421 1232 M B
431 1222 M 431 1242 L
411 1242 L
411 1222 L
431 1222 L
CF M
559 1200 M B
569 1191 M 569 1210 L
549 1210 L
549 1191 L
569 1191 L
CF M
697 1139 M B
707 1129 M 707 1149 L
687 1149 L
687 1129 L
707 1129 L
CF M
282 1253 M 421 1232 L
559 1200 L
697 1139 L
283 1042 M 284 1042 L
285 1043 L
285 1043 L
286 1043 L
287 1043 L
288 1044 L
289 1044 L
289 1044 L
290 1045 L
291 1045 L
292 1046 L
292 1047 L
293 1047 L
293 1048 L
294 1049 L
294 1049 L
295 1050 L
295 1051 L
296 1052 L
296 1053 L
296 1053 L
296 1054 L
296 1055 L
296 1056 L
296 1057 L
296 1058 L
296 1059 L
296 1059 L
296 1060 L
296 1061 L
295 1062 L
295 1063 L
294 1064 L
294 1064 L
293 1065 L
293 1066 L
292 1066 L
292 1067 L
291 1068 L
290 1068 L
289 1069 L
289 1069 L
288 1069 L
287 1070 L
286 1070 L
285 1070 L
285 1070 L
284 1070 L
283 1070 L
282 1070 L
281 1070 L
280 1070 L
279 1070 L
278 1070 L
278 1070 L
277 1069 L
276 1069 L
275 1069 L
275 1068 L
274 1068 L
273 1067 L
273 1066 L
272 1066 L
271 1065 L
271 1064 L
270 1064 L
270 1063 L
270 1062 L
269 1061 L
269 1060 L
269 1059 L
269 1059 L
268 1058 L
268 1057 L
268 1056 L
268 1055 L
269 1054 L
269 1053 L
269 1053 L
269 1052 L
270 1051 L
270 1050 L
270 1049 L
271 1049 L
271 1048 L
272 1047 L
273 1047 L
273 1046 L
274 1045 L
275 1045 L
275 1044 L
276 1044 L
277 1044 L
278 1043 L
278 1043 L
279 1043 L
280 1043 L
281 1042 L
282 1042 L
283 1042 L
421 1060 M 422 1060 L
423 1060 L
424 1060 L
425 1061 L
425 1061 L
426 1061 L
427 1061 L
428 1062 L
429 1062 L
429 1063 L
430 1063 L
431 1064 L
431 1065 L
432 1065 L
432 1066 L
433 1067 L
433 1068 L
434 1068 L
434 1069 L
434 1070 L
434 1071 L
434 1072 L
435 1073 L
435 1073 L
435 1074 L
435 1075 L
434 1076 L
434 1077 L
434 1078 L
434 1079 L
434 1080 L
433 1080 L
433 1081 L
432 1082 L
432 1083 L
431 1083 L
431 1084 L
430 1085 L
429 1085 L
429 1086 L
428 1086 L
427 1086 L
426 1087 L
425 1087 L
425 1087 L
424 1088 L
423 1088 L
422 1088 L
421 1088 L
420 1088 L
419 1088 L
418 1088 L
418 1088 L
417 1087 L
416 1087 L
415 1087 L
414 1086 L
414 1086 L
413 1086 L
412 1085 L
411 1085 L
411 1084 L
410 1083 L
410 1083 L
409 1082 L
409 1081 L
408 1080 L
408 1080 L
408 1079 L
407 1078 L
407 1077 L
407 1076 L
407 1075 L
407 1074 L
407 1073 L
407 1073 L
407 1072 L
407 1071 L
407 1070 L
408 1069 L
408 1068 L
408 1068 L
409 1067 L
409 1066 L
410 1065 L
410 1065 L
411 1064 L
411 1063 L
412 1063 L
413 1062 L
414 1062 L
414 1061 L
415 1061 L
416 1061 L
417 1061 L
418 1060 L
418 1060 L
419 1060 L
420 1060 L
421 1060 L
560 1069 M 560 1069 L
561 1070 L
562 1070 L
563 1070 L
564 1070 L
565 1071 L
565 1071 L
566 1071 L
567 1072 L
568 1072 L
568 1073 L
569 1073 L
570 1074 L
570 1075 L
571 1076 L
571 1076 L
572 1077 L
572 1078 L
572 1079 L
572 1079 L
573 1080 L
573 1081 L
573 1082 L
573 1083 L
573 1084 L
573 1085 L
573 1086 L
573 1086 L
572 1087 L
572 1088 L
572 1089 L
572 1090 L
571 1090 L
571 1091 L
570 1092 L
570 1093 L
569 1093 L
568 1094 L
568 1094 L
567 1095 L
566 1095 L
565 1096 L
565 1096 L
564 1097 L
563 1097 L
562 1097 L
561 1097 L
560 1097 L
560 1097 L
559 1097 L
558 1097 L
557 1097 L
556 1097 L
555 1097 L
554 1097 L
553 1096 L
553 1096 L
552 1095 L
551 1095 L
550 1094 L
550 1094 L
549 1093 L
548 1093 L
548 1092 L
547 1091 L
547 1090 L
547 1090 L
546 1089 L
546 1088 L
546 1087 L
545 1086 L
545 1086 L
545 1085 L
545 1084 L
545 1083 L
545 1082 L
545 1081 L
545 1080 L
546 1079 L
546 1079 L
546 1078 L
547 1077 L
547 1076 L
547 1076 L
548 1075 L
548 1074 L
549 1073 L
550 1073 L
550 1072 L
551 1072 L
552 1071 L
553 1071 L
553 1071 L
554 1070 L
555 1070 L
556 1070 L
557 1070 L
CS M
558 1069 L
559 1069 L
560 1069 L
698 1077 M 699 1077 L
700 1077 L
700 1077 L
701 1077 L
702 1077 L
703 1078 L
704 1078 L
704 1078 L
705 1079 L
706 1079 L
707 1080 L
707 1081 L
708 1081 L
708 1082 L
709 1083 L
709 1083 L
710 1084 L
710 1085 L
710 1086 L
711 1087 L
711 1088 L
711 1088 L
711 1089 L
711 1090 L
711 1091 L
711 1092 L
711 1093 L
711 1094 L
711 1094 L
710 1095 L
710 1096 L
710 1097 L
709 1098 L
709 1098 L
708 1099 L
708 1100 L
707 1100 L
707 1101 L
706 1102 L
705 1102 L
704 1103 L
704 1103 L
703 1103 L
702 1104 L
701 1104 L
700 1104 L
700 1104 L
699 1104 L
698 1104 L
697 1104 L
696 1104 L
695 1104 L
694 1104 L
693 1104 L
693 1104 L
692 1103 L
691 1103 L
690 1103 L
689 1102 L
689 1102 L
688 1101 L
687 1100 L
687 1100 L
686 1099 L
686 1098 L
685 1098 L
685 1097 L
684 1096 L
684 1095 L
684 1094 L
684 1094 L
684 1093 L
683 1092 L
683 1091 L
683 1090 L
683 1089 L
684 1088 L
684 1088 L
684 1087 L
684 1086 L
684 1085 L
685 1084 L
685 1083 L
686 1083 L
686 1082 L
687 1081 L
687 1081 L
688 1080 L
689 1079 L
689 1079 L
690 1078 L
691 1078 L
692 1078 L
693 1077 L
693 1077 L
694 1077 L
695 1077 L
696 1077 L
697 1077 L
698 1077 L
836 1080 M 837 1080 L
838 1080 L
839 1080 L
840 1080 L
840 1080 L
841 1081 L
842 1081 L
843 1081 L
844 1082 L
844 1082 L
845 1083 L
846 1084 L
846 1084 L
847 1085 L
847 1086 L
848 1086 L
848 1087 L
849 1088 L
849 1089 L
849 1090 L
849 1090 L
850 1091 L
850 1092 L
850 1093 L
850 1094 L
850 1095 L
850 1096 L
849 1097 L
849 1097 L
849 1098 L
849 1099 L
848 1100 L
848 1101 L
847 1101 L
847 1102 L
846 1103 L
846 1103 L
845 1104 L
844 1105 L
844 1105 L
843 1106 L
842 1106 L
841 1106 L
840 1107 L
840 1107 L
839 1107 L
838 1107 L
837 1107 L
836 1107 L
835 1107 L
834 1107 L
833 1107 L
833 1107 L
832 1107 L
831 1107 L
830 1106 L
829 1106 L
829 1106 L
828 1105 L
827 1105 L
826 1104 L
826 1103 L
825 1103 L
825 1102 L
824 1101 L
824 1101 L
823 1100 L
823 1099 L
823 1098 L
822 1097 L
822 1097 L
822 1096 L
822 1095 L
822 1094 L
822 1093 L
822 1092 L
822 1091 L
822 1090 L
822 1090 L
823 1089 L
823 1088 L
823 1087 L
824 1086 L
824 1086 L
825 1085 L
825 1084 L
826 1084 L
826 1083 L
827 1082 L
828 1082 L
829 1081 L
829 1081 L
830 1081 L
831 1080 L
832 1080 L
833 1080 L
833 1080 L
834 1080 L
835 1080 L
836 1080 L
975 1081 M 975 1081 L
976 1081 L
977 1081 L
978 1081 L
979 1082 L
980 1082 L
980 1082 L
981 1083 L
982 1083 L
983 1084 L
983 1084 L
984 1085 L
985 1086 L
985 1086 L
986 1087 L
986 1088 L
987 1088 L
987 1089 L
987 1090 L
988 1091 L
988 1092 L
988 1093 L
988 1093 L
988 1094 L
988 1095 L
988 1096 L
988 1097 L
988 1098 L
988 1099 L
987 1100 L
987 1100 L
987 1101 L
986 1102 L
986 1103 L
985 1103 L
985 1104 L
984 1105 L
983 1105 L
983 1106 L
982 1106 L
981 1107 L
980 1107 L
980 1108 L
979 1108 L
978 1108 L
977 1108 L
976 1109 L
975 1109 L
975 1109 L
974 1109 L
973 1109 L
972 1109 L
971 1108 L
970 1108 L
969 1108 L
968 1108 L
968 1107 L
967 1107 L
966 1106 L
965 1106 L
965 1105 L
964 1105 L
964 1104 L
963 1103 L
962 1103 L
962 1102 L
962 1101 L
961 1100 L
961 1100 L
961 1099 L
960 1098 L
960 1097 L
960 1096 L
960 1095 L
960 1094 L
960 1093 L
960 1093 L
960 1092 L
961 1091 L
961 1090 L
961 1089 L
962 1088 L
962 1088 L
962 1087 L
963 1086 L
964 1086 L
964 1085 L
965 1084 L
965 1084 L
966 1083 L
967 1083 L
968 1082 L
968 1082 L
969 1082 L
970 1081 L
971 1081 L
972 1081 L
CS M
973 1081 L
974 1081 L
975 1081 L
1113 1092 M 1114 1092 L
1115 1092 L
1115 1092 L
1116 1093 L
1117 1093 L
1118 1093 L
1119 1094 L
1119 1094 L
1120 1095 L
1121 1095 L
1122 1096 L
1122 1096 L
1123 1097 L
1123 1097 L
1124 1098 L
1124 1099 L
1125 1100 L
1125 1100 L
1126 1101 L
1126 1102 L
1126 1103 L
1126 1104 L
1126 1105 L
1126 1106 L
1126 1107 L
1126 1107 L
1126 1108 L
1126 1109 L
1126 1110 L
1126 1111 L
1125 1112 L
1125 1112 L
1124 1113 L
1124 1114 L
1123 1115 L
1123 1115 L
1122 1116 L
1122 1117 L
1121 1117 L
1120 1118 L
1119 1118 L
1119 1119 L
1118 1119 L
1117 1119 L
1116 1120 L
1115 1120 L
1115 1120 L
1114 1120 L
1113 1120 L
1112 1120 L
1111 1120 L
1110 1120 L
1109 1120 L
1108 1120 L
1108 1119 L
1107 1119 L
1106 1119 L
1105 1118 L
1104 1118 L
1104 1117 L
1103 1117 L
1102 1116 L
1102 1115 L
1101 1115 L
1101 1114 L
1100 1113 L
1100 1112 L
1099 1112 L
1099 1111 L
1099 1110 L
1099 1109 L
1099 1108 L
1098 1107 L
1098 1107 L
1098 1106 L
1098 1105 L
1099 1104 L
1099 1103 L
1099 1102 L
1099 1101 L
1099 1100 L
1100 1100 L
1100 1099 L
1101 1098 L
1101 1097 L
1102 1097 L
1102 1096 L
1103 1096 L
1104 1095 L
1104 1095 L
1105 1094 L
1106 1094 L
1107 1093 L
1108 1093 L
1108 1093 L
1109 1092 L
1110 1092 L
1111 1092 L
1112 1092 L
1113 1092 L
1251 1096 M 1252 1096 L
1253 1096 L
1254 1096 L
1255 1097 L
1255 1097 L
1256 1097 L
1257 1098 L
1258 1098 L
1259 1099 L
1259 1099 L
1260 1100 L
1261 1100 L
1261 1101 L
1262 1102 L
1262 1102 L
1263 1103 L
1263 1104 L
1264 1105 L
1264 1105 L
1264 1106 L
1264 1107 L
1265 1108 L
1265 1109 L
1265 1110 L
1265 1111 L
1265 1111 L
1265 1112 L
1264 1113 L
1264 1114 L
1264 1115 L
1264 1116 L
1263 1117 L
1263 1117 L
1262 1118 L
1262 1119 L
1261 1119 L
1261 1120 L
1260 1121 L
1259 1121 L
1259 1122 L
1258 1122 L
1257 1123 L
1256 1123 L
1255 1123 L
1255 1124 L
1254 1124 L
1253 1124 L
1252 1124 L
1251 1124 L
1250 1124 L
1249 1124 L
1248 1124 L
1248 1124 L
1247 1124 L
1246 1123 L
1245 1123 L
1244 1123 L
1244 1122 L
1243 1122 L
1242 1121 L
1241 1121 L
1241 1120 L
1240 1119 L
1240 1119 L
1239 1118 L
1239 1117 L
1238 1117 L
1238 1116 L
1238 1115 L
1237 1114 L
1237 1113 L
1237 1112 L
1237 1111 L
1237 1111 L
1237 1110 L
1237 1109 L
1237 1108 L
1237 1107 L
1237 1106 L
1238 1105 L
1238 1105 L
1238 1104 L
1239 1103 L
1239 1102 L
1240 1102 L
1240 1101 L
1241 1100 L
1241 1100 L
1242 1099 L
1243 1099 L
1244 1098 L
1244 1098 L
1245 1097 L
1246 1097 L
1247 1097 L
1248 1096 L
1248 1096 L
1249 1096 L
1250 1096 L
1251 1096 L
1389 1100 M 1390 1100 L
1391 1100 L
1392 1100 L
1393 1100 L
1394 1101 L
1395 1101 L
1395 1101 L
1396 1102 L
1397 1102 L
1398 1103 L
1398 1103 L
1399 1104 L
1400 1105 L
1400 1105 L
1401 1106 L
1401 1107 L
1401 1107 L
1402 1108 L
1402 1109 L
1402 1110 L
1403 1111 L
1403 1112 L
1403 1113 L
1403 1113 L
1403 1114 L
1403 1115 L
1403 1116 L
1403 1117 L
1402 1118 L
1402 1119 L
1402 1119 L
1401 1120 L
1401 1121 L
1401 1122 L
1400 1122 L
1400 1123 L
1399 1124 L
1398 1124 L
1398 1125 L
1397 1125 L
1396 1126 L
1395 1126 L
1395 1127 L
1394 1127 L
1393 1127 L
1392 1128 L
1391 1128 L
1390 1128 L
1389 1128 L
1389 1128 L
1388 1128 L
1387 1128 L
1386 1128 L
1385 1127 L
1384 1127 L
1383 1127 L
1383 1126 L
1382 1126 L
1381 1125 L
1380 1125 L
1380 1124 L
1379 1124 L
1378 1123 L
1378 1122 L
1377 1122 L
1377 1121 L
1376 1120 L
1376 1119 L
1376 1119 L
1376 1118 L
1375 1117 L
1375 1116 L
1375 1115 L
1375 1114 L
1375 1113 L
1375 1113 L
1375 1112 L
1375 1111 L
1376 1110 L
1376 1109 L
1376 1108 L
1376 1107 L
1377 1107 L
1377 1106 L
1378 1105 L
1378 1105 L
1379 1104 L
1380 1103 L
1380 1103 L
1381 1102 L
1382 1102 L
1383 1101 L
1383 1101 L
1384 1101 L
1385 1100 L
1386 1100 L
1387 1100 L
CS M
1388 1100 L
1389 1100 L
1389 1100 L
1528 1093 M 1529 1093 L
1530 1094 L
1530 1094 L
1531 1094 L
1532 1094 L
1533 1095 L
1534 1095 L
1534 1095 L
1535 1096 L
1536 1096 L
1537 1097 L
1537 1097 L
1538 1098 L
1538 1099 L
1539 1100 L
1539 1100 L
1540 1101 L
1540 1102 L
1541 1103 L
1541 1103 L
1541 1104 L
1541 1105 L
1541 1106 L
1541 1107 L
1541 1108 L
1541 1109 L
1541 1110 L
1541 1110 L
1541 1111 L
1541 1112 L
1540 1113 L
1540 1114 L
1539 1114 L
1539 1115 L
1538 1116 L
1538 1117 L
1537 1117 L
1537 1118 L
1536 1118 L
1535 1119 L
1534 1119 L
1534 1120 L
1533 1120 L
1532 1121 L
1531 1121 L
1530 1121 L
1530 1121 L
1529 1121 L
1528 1121 L
1527 1121 L
1526 1121 L
1525 1121 L
1524 1121 L
1523 1121 L
1523 1121 L
1522 1120 L
1521 1120 L
1520 1119 L
1520 1119 L
1519 1118 L
1518 1118 L
1517 1117 L
1517 1117 L
1516 1116 L
1516 1115 L
1515 1114 L
1515 1114 L
1514 1113 L
1514 1112 L
1514 1111 L
1514 1110 L
1514 1110 L
1513 1109 L
1513 1108 L
1513 1107 L
1513 1106 L
1514 1105 L
1514 1104 L
1514 1103 L
1514 1103 L
1514 1102 L
1515 1101 L
1515 1100 L
1516 1100 L
1516 1099 L
1517 1098 L
1517 1097 L
1518 1097 L
1519 1096 L
1520 1096 L
1520 1095 L
1521 1095 L
1522 1095 L
1523 1094 L
1523 1094 L
1524 1094 L
1525 1094 L
1526 1093 L
1527 1093 L
1528 1093 L
1666 1121 M 1667 1121 L
1668 1121 L
1669 1121 L
1670 1122 L
1670 1122 L
1671 1122 L
1672 1123 L
1673 1123 L
1673 1123 L
1674 1124 L
1675 1124 L
1676 1125 L
1676 1126 L
1677 1126 L
1677 1127 L
1678 1128 L
1678 1129 L
1679 1129 L
1679 1130 L
1679 1131 L
1679 1132 L
1679 1133 L
1680 1134 L
1680 1135 L
1680 1135 L
1680 1136 L
1679 1137 L
1679 1138 L
1679 1139 L
1679 1140 L
1679 1141 L
1678 1141 L
1678 1142 L
1677 1143 L
1677 1144 L
1676 1144 L
1676 1145 L
1675 1146 L
1674 1146 L
1673 1147 L
1673 1147 L
1672 1148 L
1671 1148 L
1670 1148 L
1670 1148 L
1669 1149 L
1668 1149 L
1667 1149 L
1666 1149 L
1665 1149 L
1664 1149 L
1663 1149 L
1663 1149 L
1662 1148 L
1661 1148 L
1660 1148 L
1659 1148 L
1659 1147 L
1658 1147 L
1657 1146 L
1656 1146 L
1656 1145 L
1655 1144 L
1655 1144 L
1654 1143 L
1654 1142 L
1653 1141 L
1653 1141 L
1652 1140 L
1652 1139 L
1652 1138 L
1652 1137 L
1652 1136 L
1652 1135 L
1652 1135 L
1652 1134 L
1652 1133 L
1652 1132 L
1652 1131 L
1652 1130 L
1653 1129 L
1653 1129 L
1654 1128 L
1654 1127 L
1655 1126 L
1655 1126 L
1656 1125 L
1656 1124 L
1657 1124 L
1658 1123 L
1659 1123 L
1659 1123 L
1660 1122 L
1661 1122 L
1662 1122 L
1663 1121 L
1663 1121 L
1664 1121 L
1665 1121 L
1666 1121 L
1804 1142 M 1805 1142 L
1806 1143 L
1807 1143 L
1808 1143 L
1809 1143 L
1810 1144 L
1810 1144 L
1811 1144 L
1812 1145 L
1813 1145 L
1813 1146 L
1814 1147 L
1815 1147 L
1815 1148 L
1816 1149 L
1816 1149 L
1817 1150 L
1817 1151 L
1817 1152 L
1817 1152 L
1818 1153 L
1818 1154 L
1818 1155 L
1818 1156 L
1818 1157 L
1818 1158 L
1818 1159 L
1818 1159 L
1817 1160 L
1817 1161 L
1817 1162 L
1817 1163 L
1816 1164 L
1816 1164 L
1815 1165 L
1815 1166 L
1814 1166 L
1813 1167 L
1813 1167 L
1812 1168 L
1811 1168 L
1810 1169 L
1810 1169 L
1809 1170 L
1808 1170 L
1807 1170 L
1806 1170 L
1805 1170 L
1804 1170 L
1804 1170 L
1803 1170 L
1802 1170 L
1801 1170 L
1800 1170 L
1799 1170 L
1798 1169 L
1798 1169 L
1797 1168 L
1796 1168 L
1795 1167 L
1795 1167 L
1794 1166 L
1793 1166 L
1793 1165 L
1792 1164 L
1792 1164 L
1791 1163 L
1791 1162 L
1791 1161 L
1791 1160 L
1790 1159 L
1790 1159 L
1790 1158 L
1790 1157 L
1790 1156 L
1790 1155 L
1790 1154 L
1790 1153 L
1791 1152 L
1791 1152 L
1791 1151 L
1791 1150 L
1792 1149 L
1792 1149 L
1793 1148 L
1793 1147 L
1794 1147 L
1795 1146 L
1795 1145 L
1796 1145 L
1797 1144 L
1798 1144 L
1798 1144 L
1799 1143 L
1800 1143 L
1801 1143 L
1802 1143 L
CS M
1803 1142 L
1804 1142 L
1804 1142 L
1943 1169 M 1944 1169 L
1945 1169 L
1945 1169 L
1946 1169 L
1947 1170 L
1948 1170 L
1949 1170 L
1949 1171 L
1950 1171 L
1951 1172 L
1952 1172 L
1952 1173 L
1953 1174 L
1953 1174 L
1954 1175 L
1954 1176 L
1955 1177 L
1955 1177 L
1955 1178 L
1956 1179 L
1956 1180 L
1956 1181 L
1956 1182 L
1956 1182 L
1956 1183 L
1956 1184 L
1956 1185 L
1956 1186 L
1956 1187 L
1955 1188 L
1955 1189 L
1955 1189 L
1954 1190 L
1954 1191 L
1953 1192 L
1953 1192 L
1952 1193 L
1952 1193 L
1951 1194 L
1950 1195 L
1949 1195 L
1949 1195 L
1948 1196 L
1947 1196 L
1946 1196 L
1945 1197 L
1945 1197 L
1944 1197 L
1943 1197 L
1942 1197 L
1941 1197 L
1940 1197 L
1939 1197 L
1938 1196 L
1938 1196 L
1937 1196 L
1936 1195 L
1935 1195 L
1934 1195 L
1934 1194 L
1933 1193 L
1932 1193 L
1932 1192 L
1931 1192 L
1931 1191 L
1930 1190 L
1930 1189 L
1929 1189 L
1929 1188 L
1929 1187 L
1929 1186 L
1928 1185 L
1928 1184 L
1928 1183 L
1928 1182 L
1928 1182 L
1928 1181 L
1929 1180 L
1929 1179 L
1929 1178 L
1929 1177 L
1930 1177 L
1930 1176 L
1931 1175 L
1931 1174 L
1932 1174 L
1932 1173 L
1933 1172 L
1934 1172 L
1934 1171 L
1935 1171 L
1936 1170 L
1937 1170 L
1938 1170 L
1938 1169 L
1939 1169 L
1940 1169 L
1941 1169 L
1942 1169 L
1943 1169 L
2081 1211 M 2082 1211 L
2083 1212 L
2084 1212 L
2085 1212 L
2085 1212 L
2086 1213 L
2087 1213 L
2088 1213 L
2088 1214 L
2089 1214 L
2090 1215 L
2091 1216 L
2091 1216 L
2092 1217 L
2092 1218 L
2093 1218 L
2093 1219 L
2094 1220 L
2094 1221 L
2094 1221 L
2094 1222 L
2094 1223 L
2095 1224 L
2095 1225 L
2095 1226 L
2095 1227 L
2094 1228 L
2094 1228 L
2094 1229 L
2094 1230 L
2094 1231 L
2093 1232 L
2093 1233 L
2092 1233 L
2092 1234 L
2091 1235 L
2091 1235 L
2090 1236 L
2089 1236 L
2088 1237 L
2088 1237 L
2087 1238 L
2086 1238 L
2085 1239 L
2085 1239 L
2084 1239 L
2083 1239 L
2082 1239 L
2081 1239 L
2080 1239 L
2079 1239 L
2078 1239 L
2078 1239 L
2077 1239 L
2076 1239 L
2075 1238 L
2074 1238 L
2074 1237 L
2073 1237 L
2072 1236 L
2071 1236 L
2071 1235 L
2070 1235 L
2070 1234 L
2069 1233 L
2069 1233 L
2068 1232 L
2068 1231 L
2067 1230 L
2067 1229 L
2067 1228 L
2067 1228 L
2067 1227 L
2067 1226 L
2067 1225 L
2067 1224 L
2067 1223 L
2067 1222 L
2067 1221 L
2067 1221 L
2068 1220 L
2068 1219 L
2069 1218 L
2069 1218 L
2070 1217 L
2070 1216 L
2071 1216 L
2071 1215 L
2072 1214 L
2073 1214 L
2074 1213 L
2074 1213 L
2075 1213 L
2076 1212 L
2077 1212 L
2078 1212 L
2078 1212 L
2079 1211 L
2080 1211 L
2081 1211 L
CS [6 12] 0 setdash M
282 1057 M 421 1074 L
559 1083 L
697 1091 L
836 1094 L
974 1095 L
1112 1106 L
1251 1110 L
1389 1114 L
1527 1107 L
1666 1135 L
1804 1156 L
1942 1183 L
2081 1225 L
CS [] 0 setdash M
1105 100 M 1126 151 M 1126 100 L
1129 151 M 1129 100 L
1112 151 M 1110 136 L
1110 151 L
1146 151 L
1146 136 L
1143 151 L
1119 100 M 1136 100 L
1216 160 M 1211 156 L
1206 148 L
1201 139 L
1199 127 L
1199 117 L
1201 105 L
1206 95 L
1211 88 L
1216 83 L
1211 156 M 1206 146 L
1203 139 L
1201 127 L
1201 117 L
1203 105 L
1206 98 L
1211 88 L
1235 151 M 1235 100 L
1237 151 M 1252 108 L
1235 151 M 1252 100 L
1268 151 M 1252 100 L
1268 151 M 1268 100 L
1271 151 M 1271 100 L
1228 151 M 1237 151 L
1268 151 M 1278 151 L
1228 100 M 1242 100 L
1261 100 M 1278 100 L
1293 120 M 1321 120 L
1321 124 L
1319 129 L
1317 132 L
1312 134 L
1305 134 L
1297 132 L
1293 127 L
1290 120 L
1290 115 L
1293 108 L
1297 103 L
1305 100 L
1309 100 L
1317 103 L
1321 108 L
1319 120 M 1319 127 L
1317 132 L
1305 134 M 1300 132 L
1295 127 L
1293 120 L
1293 115 L
1295 108 L
1300 103 L
1305 100 L
1336 151 M 1353 100 L
1338 151 M 1353 108 L
1370 151 M 1353 100 L
1331 151 M 1345 151 L
1360 151 M 1374 151 L
1384 160 M 1389 156 L
1394 148 L
1398 139 L
1401 127 L
1401 117 L
1398 105 L
1394 95 L
1389 88 L
1384 83 L
1389 156 M 1394 146 L
1396 139 L
1398 127 L
1398 117 L
1396 105 L
1394 98 L
1389 88 L
CS [6 12] 0 setdash M
CS [] 0 setdash M
146 1188 M 117 1219 M 112 1222 L
122 1222 L
117 1219 L
115 1217 L
CS M
112 1212 L
112 1202 L
115 1197 L
117 1195 L
122 1195 L
124 1197 L
127 1202 L
132 1214 L
134 1219 L
136 1222 L
120 1195 M 122 1197 L
124 1202 L
129 1214 L
132 1219 L
134 1222 L
141 1222 L
144 1219 L
146 1214 L
146 1205 L
144 1200 L
141 1197 L
136 1195 L
146 1195 L
141 1197 L
86 1277 M 163 1234 L
134 1291 M 141 1294 L
144 1296 L
146 1301 L
146 1306 L
144 1313 L
136 1318 L
129 1320 L
122 1320 L
117 1318 L
115 1315 L
112 1311 L
112 1306 L
115 1299 L
122 1294 L
129 1291 L
163 1282 L
146 1306 M 144 1311 L
136 1315 L
129 1318 L
120 1318 L
115 1315 L
112 1306 M 115 1301 L
122 1296 L
129 1294 L
163 1284 L
CS [6 12] 0 setdash M
1527 1879 M CS [] 0 setdash M
1361 1854 M 1378 1888 M 1364 1837 L
1381 1888 M 1366 1837 L
1378 1881 M 1376 1866 L
1376 1859 L
1381 1854 L
1385 1854 L
1390 1857 L
1395 1861 L
1400 1869 L
1405 1888 M 1397 1861 L
1397 1857 L
1400 1854 L
1407 1854 L
1412 1859 L
1414 1864 L
1407 1888 M 1400 1861 L
1400 1857 L
1402 1854 L
1426 1883 M 1470 1883 L
1426 1869 M 1470 1869 L
1501 1905 M 1494 1902 L
1489 1895 L
1486 1883 L
1486 1876 L
1489 1864 L
1494 1857 L
1501 1854 L
1506 1854 L
1513 1857 L
1518 1864 L
1520 1876 L
1520 1883 L
1518 1895 L
1513 1902 L
1506 1905 L
1501 1905 L
1496 1902 L
1494 1900 L
1491 1895 L
1489 1883 L
1489 1876 L
1491 1864 L
1494 1859 L
1496 1857 L
1501 1854 L
1506 1854 M 1511 1857 L
1513 1859 L
1515 1864 L
1518 1876 L
1518 1883 L
1515 1895 L
1513 1900 L
1511 1902 L
1506 1905 L
CS [6 12] 0 setdash M
1527 1799 M CS [] 0 setdash M
1431 1774 M 1441 1815 M 1443 1812 L
1441 1810 L
1438 1812 L
1438 1815 L
1441 1820 L
1443 1822 L
1450 1824 L
1460 1824 L
1467 1822 L
1470 1820 L
1472 1815 L
1472 1810 L
1470 1805 L
1462 1800 L
1450 1796 L
1446 1793 L
1441 1788 L
1438 1781 L
1438 1774 L
1460 1824 M 1465 1822 L
1467 1820 L
1470 1815 L
1470 1810 L
1467 1805 L
1460 1800 L
1450 1796 L
1438 1779 M 1441 1781 L
1446 1781 L
1458 1776 L
1465 1776 L
1470 1779 L
1472 1781 L
1446 1781 M 1458 1774 L
1467 1774 L
1470 1776 L
1472 1781 L
1472 1786 L
1501 1824 M 1494 1822 L
1489 1815 L
1486 1803 L
1486 1796 L
1489 1783 L
1494 1776 L
1501 1774 L
1506 1774 L
1513 1776 L
1518 1783 L
1520 1796 L
1520 1803 L
1518 1815 L
1513 1822 L
1506 1824 L
1501 1824 L
1496 1822 L
1494 1820 L
1491 1815 L
1489 1803 L
1489 1796 L
1491 1783 L
1494 1779 L
1496 1776 L
1501 1774 L
1506 1774 M 1511 1776 L
1513 1779 L
1515 1783 L
1518 1796 L
1518 1803 L
1515 1815 L
1513 1820 L
1511 1822 L
1506 1824 L
CS [6 12] 0 setdash M
1527 1699 M CS [] 0 setdash M
1431 1674 M 1460 1719 M 1460 1674 L
1462 1724 M 1462 1674 L
1462 1724 M 1436 1688 L
1474 1688 L
1453 1674 M 1470 1674 L
1501 1724 M 1494 1722 L
1489 1714 L
1486 1702 L
1486 1695 L
1489 1683 L
1494 1676 L
1501 1674 L
1506 1674 L
1513 1676 L
1518 1683 L
1520 1695 L
1520 1702 L
1518 1714 L
1513 1722 L
1506 1724 L
1501 1724 L
1496 1722 L
1494 1719 L
1491 1714 L
1489 1702 L
1489 1695 L
1491 1683 L
1494 1678 L
1496 1676 L
1501 1674 L
1506 1674 M 1511 1676 L
1513 1678 L
1515 1683 L
1518 1695 L
1518 1702 L
1515 1714 L
1513 1719 L
1511 1722 L
1506 1724 L
CS [6 12] 0 setdash M
1527 1599 M CS [] 0 setdash M
1431 1573 M 1467 1617 M 1465 1614 L
1467 1612 L
1470 1614 L
1470 1617 L
1467 1621 L
1462 1624 L
1455 1624 L
1448 1621 L
1443 1617 L
1441 1612 L
1438 1602 L
1438 1588 L
1441 1580 L
1446 1576 L
1453 1573 L
1458 1573 L
1465 1576 L
1470 1580 L
1472 1588 L
1472 1590 L
1470 1597 L
1465 1602 L
1458 1605 L
1455 1605 L
1448 1602 L
1443 1597 L
1441 1590 L
1455 1624 M 1450 1621 L
1446 1617 L
1443 1612 L
1441 1602 L
1441 1588 L
1443 1580 L
1448 1576 L
1453 1573 L
1458 1573 M 1462 1576 L
1467 1580 L
1470 1588 L
1470 1590 L
1467 1597 L
1462 1602 L
1458 1605 L
1501 1624 M 1494 1621 L
1489 1614 L
1486 1602 L
1486 1595 L
1489 1583 L
1494 1576 L
1501 1573 L
1506 1573 L
1513 1576 L
1518 1583 L
1520 1595 L
1520 1602 L
1518 1614 L
1513 1621 L
1506 1624 L
1501 1624 L
1496 1621 L
1494 1619 L
1491 1614 L
1489 1602 L
1489 1595 L
1491 1583 L
1494 1578 L
1496 1576 L
1501 1573 L
1506 1573 M 1511 1576 L
1513 1578 L
1515 1583 L
1518 1595 L
1518 1602 L
1515 1614 L
1513 1619 L
1511 1621 L
1506 1624 L
CS [6 12] 0 setdash M
1527 1518 M CS [] 0 setdash M
1431 1493 M 1450 1544 M 1443 1541 L
1441 1536 L
1441 1529 L
1443 1524 L
1450 1522 L
1460 1522 L
1467 1524 L
1470 1529 L
1470 1536 L
1467 1541 L
CS M
1460 1544 L
1450 1544 L
1446 1541 L
1443 1536 L
1443 1529 L
1446 1524 L
1450 1522 L
1460 1522 M 1465 1524 L
1467 1529 L
1467 1536 L
1465 1541 L
1460 1544 L
1450 1522 M 1443 1520 L
1441 1517 L
1438 1512 L
1438 1503 L
1441 1498 L
1443 1495 L
1450 1493 L
1460 1493 L
1467 1495 L
1470 1498 L
1472 1503 L
1472 1512 L
1470 1517 L
1467 1520 L
1460 1522 L
1450 1522 M 1446 1520 L
1443 1517 L
1441 1512 L
1441 1503 L
1443 1498 L
1446 1495 L
1450 1493 L
1460 1493 M 1465 1495 L
1467 1498 L
1470 1503 L
1470 1512 L
1467 1517 L
1465 1520 L
1460 1522 L
1501 1544 M 1494 1541 L
1489 1534 L
1486 1522 L
1486 1515 L
1489 1503 L
1494 1495 L
1501 1493 L
1506 1493 L
1513 1495 L
1518 1503 L
1520 1515 L
1520 1522 L
1518 1534 L
1513 1541 L
1506 1544 L
1501 1544 L
1496 1541 L
1494 1539 L
1491 1534 L
1489 1522 L
1489 1515 L
1491 1503 L
1494 1498 L
1496 1495 L
1501 1493 L
1506 1493 M 1511 1495 L
1513 1498 L
1515 1503 L
1518 1515 L
1518 1522 L
1515 1534 L
1513 1539 L
1511 1541 L
1506 1544 L
CS [6 12] 0 setdash M
1514 1438 M CS [] 0 setdash M
1417 1413 M 1456 1447 M 1453 1439 L
1449 1435 L
1441 1432 L
1439 1432 L
1432 1435 L
1427 1439 L
1424 1447 L
1424 1449 L
1427 1456 L
1432 1461 L
1439 1463 L
1444 1463 L
1451 1461 L
1456 1456 L
1458 1449 L
1458 1435 L
1456 1425 L
1453 1420 L
1449 1415 L
1441 1413 L
1434 1413 L
1429 1415 L
1427 1420 L
1427 1423 L
1429 1425 L
1432 1423 L
1429 1420 L
1439 1432 M 1434 1435 L
1429 1439 L
1427 1447 L
1427 1449 L
1429 1456 L
1434 1461 L
1439 1463 L
1444 1463 M 1449 1461 L
1453 1456 L
1456 1449 L
1456 1435 L
1453 1425 L
1451 1420 L
1446 1415 L
1441 1413 L
1487 1463 M 1480 1461 L
1475 1454 L
1473 1442 L
1473 1435 L
1475 1423 L
1480 1415 L
1487 1413 L
1492 1413 L
1499 1415 L
1504 1423 L
1506 1435 L
1506 1442 L
1504 1454 L
1499 1461 L
1492 1463 L
1487 1463 L
1482 1461 L
1480 1459 L
1477 1454 L
1475 1442 L
1475 1435 L
1477 1423 L
1480 1418 L
1482 1415 L
1487 1413 L
1492 1413 M 1497 1415 L
1499 1418 L
1501 1423 L
1504 1435 L
1504 1442 L
1501 1454 L
1499 1459 L
1497 1461 L
1492 1463 L
CS [6 12] 0 setdash M
1486 1378 M CS [] 0 setdash M
1341 1353 M 1356 1394 M 1361 1396 L
1368 1403 L
1368 1353 L
1366 1401 M 1366 1353 L
1356 1353 M 1378 1353 L
1411 1403 M 1404 1401 L
1399 1394 L
1397 1382 L
1397 1374 L
1399 1362 L
1404 1355 L
1411 1353 L
1416 1353 L
1423 1355 L
1428 1362 L
1431 1374 L
1431 1382 L
1428 1394 L
1423 1401 L
1416 1403 L
1411 1403 L
1406 1401 L
1404 1398 L
1402 1394 L
1399 1382 L
1399 1374 L
1402 1362 L
1404 1357 L
1406 1355 L
1411 1353 L
1416 1353 M 1421 1355 L
1423 1357 L
1426 1362 L
1428 1374 L
1428 1382 L
1426 1394 L
1423 1398 L
1421 1401 L
1416 1403 L
1459 1403 M 1452 1401 L
1447 1394 L
1445 1382 L
1445 1374 L
1447 1362 L
1452 1355 L
1459 1353 L
1464 1353 L
1471 1355 L
1476 1362 L
1479 1374 L
1479 1382 L
1476 1394 L
1471 1401 L
1464 1403 L
1459 1403 L
1455 1401 L
1452 1398 L
1450 1394 L
1447 1382 L
1447 1374 L
1450 1362 L
1452 1357 L
1455 1355 L
1459 1353 L
1464 1353 M 1469 1355 L
1471 1357 L
1474 1362 L
1476 1374 L
1476 1382 L
1474 1394 L
1471 1398 L
1469 1401 L
1464 1403 L
CS [6 12] 0 setdash M
1389 1298 M CS [] 0 setdash M
1245 1272 M 1259 1313 M 1264 1316 L
1271 1323 L
1271 1272 L
1269 1321 M 1269 1272 L
1259 1272 M 1281 1272 L
1307 1313 M 1312 1316 L
1319 1323 L
1319 1272 L
1317 1321 M 1317 1272 L
1307 1272 M 1329 1272 L
1363 1323 M 1355 1321 L
1351 1313 L
1348 1301 L
1348 1294 L
1351 1282 L
1355 1275 L
1363 1272 L
1367 1272 L
1375 1275 L
1379 1282 L
1382 1294 L
1382 1301 L
1379 1313 L
1375 1321 L
1367 1323 L
1363 1323 L
1358 1321 L
1355 1318 L
1353 1313 L
1351 1301 L
1351 1294 L
1353 1282 L
1355 1277 L
1358 1275 L
1363 1272 L
1367 1272 M 1372 1275 L
1375 1277 L
1377 1282 L
1379 1294 L
1379 1301 L
1377 1313 L
1375 1318 L
1372 1321 L
1367 1323 L
CS [6 12] 0 setdash M
1182 1218 M CS [] 0 setdash M
1037 1192 M 1052 1233 M 1056 1236 L
1064 1243 L
1064 1192 L
1061 1240 M 1061 1192 L
1052 1192 M 1073 1192 L
1095 1233 M 1097 1231 L
1095 1228 L
1092 1231 L
1092 1233 L
1095 1238 L
1097 1240 L
1104 1243 L
1114 1243 L
1121 1240 L
1124 1238 L
1126 1233 L
1126 1228 L
1124 1224 L
1117 1219 L
1104 1214 L
1100 1211 L
1095 1207 L
1092 1199 L
1092 1192 L
1114 1243 M 1119 1240 L
1121 1238 L
1124 1233 L
1124 1228 L
1121 1224 L
1114 1219 L
1104 1214 L
1092 1197 M 1095 1199 L
CS M
1100 1199 L
1112 1195 L
1119 1195 L
1124 1197 L
1126 1199 L
1100 1199 M 1112 1192 L
1121 1192 L
1124 1195 L
1126 1199 L
1126 1204 L
1155 1243 M 1148 1240 L
1143 1233 L
1141 1221 L
1141 1214 L
1143 1202 L
1148 1195 L
1155 1192 L
1160 1192 L
1167 1195 L
1172 1202 L
1174 1214 L
1174 1221 L
1172 1233 L
1167 1240 L
1160 1243 L
1155 1243 L
1150 1240 L
1148 1238 L
1145 1233 L
1143 1221 L
1143 1214 L
1145 1202 L
1148 1197 L
1150 1195 L
1155 1192 L
1160 1192 M 1165 1195 L
1167 1197 L
1169 1202 L
1172 1214 L
1172 1221 L
1169 1233 L
1167 1238 L
1165 1240 L
1160 1243 L
CS [6 12] 0 setdash M
836 1177 M CS [] 0 setdash M
691 1152 M 706 1193 M 711 1195 L
718 1203 L
718 1152 L
715 1200 M 715 1152 L
706 1152 M 727 1152 L
749 1193 M 751 1191 L
749 1188 L
747 1191 L
747 1193 L
749 1198 L
751 1200 L
759 1203 L
768 1203 L
776 1200 L
778 1195 L
778 1188 L
776 1183 L
768 1181 L
761 1181 L
768 1203 M 773 1200 L
776 1195 L
776 1188 L
773 1183 L
768 1181 L
773 1179 L
778 1174 L
780 1169 L
780 1162 L
778 1157 L
776 1154 L
768 1152 L
759 1152 L
751 1154 L
749 1157 L
747 1162 L
747 1164 L
749 1166 L
751 1164 L
749 1162 L
776 1176 M 778 1169 L
778 1162 L
776 1157 L
773 1154 L
768 1152 L
809 1203 M 802 1200 L
797 1193 L
795 1181 L
795 1174 L
797 1162 L
802 1154 L
809 1152 L
814 1152 L
821 1154 L
826 1162 L
829 1174 L
829 1181 L
826 1193 L
821 1200 L
814 1203 L
809 1203 L
804 1200 L
802 1198 L
800 1193 L
797 1181 L
797 1174 L
800 1162 L
802 1157 L
804 1154 L
809 1152 L
814 1152 M 819 1154 L
821 1157 L
824 1162 L
826 1174 L
826 1181 L
824 1193 L
821 1198 L
819 1200 L
814 1203 L
CS [6 12] 0 setdash M
stroke
grestore
showpage
end


 20 dict begin
72 300 div dup scale
1 setlinejoin 0 setlinecap
/Helvetica findfont 55 scalefont setfont
/B { stroke newpath } def /F { moveto 0 setlinecap} def
/C { CS M 1 1 3 { pop 3 1 roll 255 div } for SET_COLOUR } def
/CS { currentpoint stroke } def
/CF { currentpoint fill } def
/L { lineto } def /M { moveto } def
/P { moveto 0 1 rlineto stroke } def
/T { 1 setlinecap show } def
errordict /nocurrentpoint { pop 0 0 M currentpoint } put
/SET_COLOUR { pop pop pop } def
 80 600 translate
gsave
CS [] 0 setdash M
CS M 2 setlinewidth
/P { moveto 0 2.05 rlineto stroke } def
 0 0 0 C
CS [] 0 setdash M
255 255 M 2261 255 L
282 255 M 282 280 L
421 255 M 421 280 L
559 255 M 559 306 L
697 255 M 697 280 L
836 255 M 836 280 L
974 255 M 974 280 L
1112 255 M 1112 280 L
1251 255 M 1251 306 L
1389 255 M 1389 280 L
1527 255 M 1527 280 L
1666 255 M 1666 280 L
1804 255 M 1804 280 L
1942 255 M 1942 306 L
2081 255 M 2081 280 L
2219 255 M 2219 280 L
487 182 M 501 223 M 506 226 L
513 233 L
513 182 L
511 230 M 511 182 L
501 182 M 523 182 L
557 233 M 549 230 L
545 223 L
542 211 L
542 204 L
545 192 L
549 185 L
557 182 L
561 182 L
569 185 L
574 192 L
576 204 L
576 211 L
574 223 L
569 230 L
561 233 L
557 233 L
552 230 L
549 228 L
547 223 L
545 211 L
545 204 L
547 192 L
549 187 L
552 185 L
557 182 L
561 182 M 566 185 L
569 187 L
571 192 L
574 204 L
574 211 L
571 223 L
569 228 L
566 230 L
561 233 L
605 233 M 598 230 L
593 223 L
590 211 L
590 204 L
593 192 L
598 185 L
605 182 L
610 182 L
617 185 L
622 192 L
624 204 L
624 211 L
622 223 L
617 230 L
610 233 L
605 233 L
600 230 L
598 228 L
595 223 L
593 211 L
593 204 L
595 192 L
598 187 L
600 185 L
605 182 L
610 182 M 614 185 L
617 187 L
619 192 L
622 204 L
622 211 L
619 223 L
617 228 L
614 230 L
610 233 L
1179 182 M 1193 223 M 1198 226 L
1205 233 L
1205 182 L
1203 230 M 1203 182 L
1193 182 M 1215 182 L
1239 233 M 1234 209 L
1239 214 L
1246 216 L
1253 216 L
1260 214 L
1265 209 L
1268 202 L
1268 197 L
1265 190 L
1260 185 L
1253 182 L
1246 182 L
1239 185 L
1236 187 L
1234 192 L
1234 194 L
1236 197 L
1239 194 L
1236 192 L
1253 216 M 1258 214 L
1263 209 L
1265 202 L
1265 197 L
1263 190 L
1258 185 L
1253 182 L
1239 233 M 1263 233 L
1239 230 M 1251 230 L
1263 233 L
1296 233 M 1289 230 L
1284 223 L
1282 211 L
1282 204 L
1284 192 L
1289 185 L
1296 182 L
1301 182 L
1309 185 L
1313 192 L
1316 204 L
1316 211 L
1313 223 L
1309 230 L
1301 233 L
1296 233 L
1292 230 L
1289 228 L
1287 223 L
1284 211 L
1284 204 L
1287 192 L
1289 187 L
1292 185 L
1296 182 L
1301 182 M 1306 185 L
1309 187 L
1311 192 L
1313 204 L
1313 211 L
1311 223 L
1309 228 L
1306 230 L
1301 233 L
1870 182 M 1880 223 M 1882 221 L
1880 218 L
1877 221 L
1877 223 L
1880 228 L
1882 230 L
1889 233 L
1899 233 L
1906 230 L
1909 228 L
1911 223 L
1911 218 L
1909 214 L
1901 209 L
1889 204 L
1885 202 L
1880 197 L
1877 190 L
1877 182 L
1899 233 M 1904 230 L
1906 228 L
1909 223 L
1909 218 L
1906 214 L
1899 209 L
1889 204 L
1877 187 M 1880 190 L
1885 190 L
1897 185 L
1904 185 L
1909 187 L
1911 190 L
1885 190 M 1897 182 L
1906 182 L
1909 185 L
1911 190 L
1911 194 L
1940 233 M 1933 230 L
1928 223 L
1925 211 L
1925 204 L
1928 192 L
1933 185 L
1940 182 L
1945 182 L
1952 185 L
1957 192 L
1959 204 L
1959 211 L
1957 223 L
1952 230 L
1945 233 L
1940 233 L
1935 230 L
1933 228 L
1930 223 L
1928 211 L
1928 204 L
1930 192 L
1933 187 L
1935 185 L
1940 182 L
1945 182 M 1950 185 L
1952 187 L
1954 192 L
1957 204 L
1957 211 L
1954 223 L
1952 228 L
1950 230 L
1945 233 L
1988 233 M 1981 230 L
1976 223 L
1974 211 L
1974 204 L
1976 192 L
1981 185 L
1988 182 L
1993 182 L
2000 185 L
2005 192 L
2007 204 L
2007 211 L
2005 223 L
2000 230 L
1993 233 L
1988 233 L
1983 230 L
1981 228 L
1978 223 L
1976 211 L
1976 204 L
1978 192 L
1981 187 L
1983 185 L
1988 182 L
1993 182 M 1998 185 L
2000 187 L
2002 192 L
2005 204 L
2005 211 L
2002 223 L
2000 228 L
1998 230 L
1993 233 L
255 2261 M 2261 2261 L
282 2261 M 282 2235 L
421 2261 M 421 2235 L
559 2261 M 559 2209 L
697 2261 M 697 2235 L
836 2261 M 836 2235 L
974 2261 M 974 2235 L
1112 2261 M 1112 2235 L
1251 2261 M 1251 2209 L
1389 2261 M 1389 2235 L
1527 2261 M 1527 2235 L
1666 2261 M 1666 2235 L
1804 2261 M 1804 2235 L
1942 2261 M 1942 2209 L
2081 2261 M 2081 2235 L
2219 2261 M 2219 2235 L
255 255 M 255 2261 L
255 255 M 306 255 L
255 335 M 280 335 L
255 415 M 280 415 L
255 495 M 280 495 L
255 576 M 280 576 L
255 656 M 306 656 L
255 736 M 280 736 L
255 816 M 280 816 L
255 897 M 280 897 L
255 977 M 280 977 L
255 1057 M 306 1057 L
255 1137 M 280 1137 L
255 1218 M 280 1218 L
255 1298 M 280 1298 L
255 1378 M 280 1378 L
255 1458 M 306 1458 L
255 1538 M 280 1538 L
255 1619 M 280 1619 L
255 1699 M 280 1699 L
255 1779 M 280 1779 L
255 1859 M 306 1859 L
CS M
255 1940 M 280 1940 L
255 2020 M 280 2020 L
255 2100 M 280 2100 L
255 2180 M 280 2180 L
255 2261 M 306 2261 L
185 229 M 206 280 M 199 278 L
194 270 L
192 258 L
192 251 L
194 239 L
199 232 L
206 229 L
211 229 L
218 232 L
223 239 L
226 251 L
226 258 L
223 270 L
218 278 L
211 280 L
206 280 L
202 278 L
199 275 L
197 270 L
194 258 L
194 251 L
197 239 L
199 234 L
202 232 L
206 229 L
211 229 M 216 232 L
218 234 L
221 239 L
223 251 L
223 258 L
221 270 L
218 275 L
216 278 L
211 280 L
185 631 M 199 672 M 204 674 L
211 681 L
211 631 L
209 679 M 209 631 L
199 631 M 221 631 L
185 1032 M 194 1073 M 197 1070 L
194 1068 L
192 1070 L
192 1073 L
194 1078 L
197 1080 L
204 1082 L
214 1082 L
221 1080 L
223 1078 L
226 1073 L
226 1068 L
223 1063 L
216 1058 L
204 1053 L
199 1051 L
194 1046 L
192 1039 L
192 1032 L
214 1082 M 218 1080 L
221 1078 L
223 1073 L
223 1068 L
221 1063 L
214 1058 L
204 1053 L
192 1037 M 194 1039 L
199 1039 L
211 1034 L
218 1034 L
223 1037 L
226 1039 L
199 1039 M 211 1032 L
221 1032 L
223 1034 L
226 1039 L
226 1044 L
185 1433 M 194 1474 M 197 1471 L
194 1469 L
192 1471 L
192 1474 L
194 1479 L
197 1481 L
204 1483 L
214 1483 L
221 1481 L
223 1476 L
223 1469 L
221 1464 L
214 1462 L
206 1462 L
214 1483 M 218 1481 L
221 1476 L
221 1469 L
218 1464 L
214 1462 L
218 1459 L
223 1455 L
226 1450 L
226 1443 L
223 1438 L
221 1435 L
214 1433 L
204 1433 L
197 1435 L
194 1438 L
192 1443 L
192 1445 L
194 1447 L
197 1445 L
194 1443 L
221 1457 M 223 1450 L
223 1443 L
221 1438 L
218 1435 L
214 1433 L
185 1834 M 214 1880 M 214 1834 L
216 1885 M 216 1834 L
216 1885 M 190 1849 L
228 1849 L
206 1834 M 223 1834 L
185 2235 M 197 2286 M 192 2262 L
197 2267 L
204 2269 L
211 2269 L
218 2267 L
223 2262 L
226 2254 L
226 2250 L
223 2242 L
218 2238 L
211 2235 L
204 2235 L
197 2238 L
194 2240 L
192 2245 L
192 2247 L
194 2250 L
197 2247 L
194 2245 L
211 2269 M 216 2267 L
221 2262 L
223 2254 L
223 2250 L
221 2242 L
216 2238 L
211 2235 L
197 2286 M 221 2286 L
197 2283 M 209 2283 L
221 2286 L
2261 255 M 2261 2261 L
2261 255 M 2209 255 L
2261 335 M 2235 335 L
2261 415 M 2235 415 L
2261 495 M 2235 495 L
2261 576 M 2235 576 L
2261 656 M 2209 656 L
2261 736 M 2235 736 L
2261 816 M 2235 816 L
2261 897 M 2235 897 L
2261 977 M 2235 977 L
2261 1057 M 2209 1057 L
2261 1137 M 2235 1137 L
2261 1218 M 2235 1218 L
2261 1298 M 2235 1298 L
2261 1378 M 2235 1378 L
2261 1458 M 2209 1458 L
2261 1538 M 2235 1538 L
2261 1619 M 2235 1619 L
2261 1699 M 2235 1699 L
2261 1779 M 2235 1779 L
2261 1859 M 2209 1859 L
2261 1940 M 2235 1940 L
2261 2020 M 2235 2020 L
2261 2100 M 2235 2100 L
2261 2180 M 2235 2180 L
2261 2261 M 2209 2261 L
CS [] 0 setdash M
282 2191 M B
292 2181 M 292 2200 L
273 2200 L
273 2181 L
292 2181 L
CF M
421 2124 M B
431 2114 M 431 2134 L
411 2134 L
411 2114 L
431 2114 L
CF M
559 2072 M B
569 2062 M 569 2082 L
549 2082 L
549 2062 L
569 2062 L
CF M
697 2029 M B
707 2019 M 707 2039 L
687 2039 L
687 2020 L
707 2020 L
CF M
836 1995 M B
846 1985 M 846 2005 L
826 2005 L
826 1985 L
846 1985 L
CF M
974 1966 M B
984 1956 M 984 1976 L
964 1976 L
964 1956 L
984 1956 L
CF M
1112 1941 M B
1122 1931 M 1122 1951 L
1102 1951 L
1102 1931 L
1122 1931 L
CF M
1251 1920 M B
1261 1910 M 1261 1930 L
1241 1930 L
1241 1910 L
1261 1910 L
CF M
1389 1902 M B
1399 1892 M 1399 1912 L
1379 1912 L
1379 1892 L
1399 1892 L
CF M
1527 1887 M B
1537 1877 M 1537 1897 L
1517 1897 L
1517 1877 L
1537 1877 L
CF M
1666 1873 M B
1676 1863 M 1676 1883 L
1656 1883 L
1656 1863 L
1676 1863 L
CF M
1804 1861 M B
1814 1851 M 1814 1870 L
1794 1870 L
1794 1851 L
1814 1851 L
CF M
1942 1850 M B
1952 1840 M 1952 1860 L
1932 1860 L
1932 1840 L
1952 1840 L
CF M
2081 1840 M B
2091 1830 M 2091 1850 L
2071 1850 L
2071 1830 L
2091 1830 L
CF M
2219 1831 M B
2229 1821 M 2229 1841 L
2209 1841 L
2209 1821 L
2229 1821 L
CF M
282 2191 M 421 2124 L
559 2072 L
697 2029 L
836 1995 L
974 1966 L
1112 1941 L
1251 1920 L
1389 1902 L
1527 1887 L
1666 1873 L
1804 1861 L
1942 1850 L
2081 1840 L
2219 1831 L
282 2081 M B
292 2071 M 292 2090 L
273 2090 L
273 2071 L
292 2071 L
CF M
421 2024 M B
431 2014 M 431 2034 L
411 2034 L
411 2014 L
431 2014 L
CF M
559 1980 M B
569 1970 M 569 1990 L
549 1990 L
549 1970 L
569 1970 L
CF M
697 1945 M B
707 1935 M 707 1955 L
687 1955 L
687 1935 L
707 1935 L
CF M
836 1916 M B
846 1906 M 846 1926 L
826 1926 L
826 1906 L
846 1906 L
CF M
974 1892 M B
984 1882 M 984 1902 L
964 1902 L
964 1882 L
984 1882 L
CF M
1112 1872 M B
1122 1862 M 1122 1881 L
1102 1881 L
1102 1862 L
1122 1862 L
CF M
1251 1854 M B
1261 1844 M 1261 1864 L
1241 1864 L
1241 1844 L
1261 1844 L
CF M
1389 1839 M B
1399 1830 M 1399 1849 L
1379 1849 L
1379 1830 L
1399 1830 L
CF M
1527 1827 M B
1537 1817 M 1537 1837 L
1517 1837 L
1517 1817 L
1537 1817 L
CF M
1666 1815 M B
1676 1806 M 1676 1825 L
1656 1825 L
1656 1806 L
1676 1806 L
CF M
1804 1806 M B
1814 1796 M 1814 1816 L
1794 1816 L
1794 1796 L
1814 1796 L
CF M
1942 1797 M B
1952 1787 M 1952 1807 L
1932 1807 L
1932 1787 L
1952 1787 L
CF M
2081 1789 M B
2091 1779 M 2091 1799 L
2071 1799 L
2071 1779 L
2091 1779 L
CF M
2219 1782 M B
2229 1772 M 2229 1792 L
2209 1792 L
2209 1772 L
2229 1772 L
CF M
282 2081 M 421 2024 L
559 1980 L
697 1945 L
836 1916 L
974 1892 L
1112 1872 L
1251 1854 L
1389 1839 L
1527 1827 L
1666 1815 L
1804 1806 L
1942 1797 L
2081 1789 L
2219 1782 L
282 1967 M B
292 1957 M 292 1977 L
273 1977 L
273 1957 L
292 1957 L
CF M
421 1921 M B
431 1911 M 431 1931 L
411 1931 L
411 1911 L
431 1911 L
CF M
559 1885 M B
569 1875 M 569 1895 L
549 1895 L
549 1875 L
569 1875 L
CF M
697 1857 M B
707 1847 M 707 1867 L
687 1867 L
687 1847 L
707 1847 L
CF M
836 1834 M B
846 1824 M 846 1844 L
826 1844 L
826 1824 L
846 1824 L
CF M
974 1815 M B
984 1805 M 984 1824 L
964 1824 L
964 1805 L
984 1805 L
CF M
1112 1799 M B
1122 1789 M 1122 1808 L
1102 1808 L
1102 1789 L
1122 1789 L
CF M
1251 1785 M B
1261 1775 M 1261 1795 L
1241 1795 L
1241 1775 L
1261 1775 L
CF M
1389 1774 M B
1399 1764 M 1399 1784 L
1379 1784 L
1379 1764 L
1399 1764 L
CF M
1527 1764 M B
1537 1754 M 1537 1774 L
1517 1774 L
1517 1754 L
1537 1754 L
CF M
1666 1756 M B
1676 1746 M 1676 1765 L
1656 1765 L
1656 1746 L
1676 1746 L
CF M
1804 1748 M B
1814 1738 M 1814 1758 L
1794 1758 L
1794 1738 L
1814 1738 L
CF M
1942 1742 M B
1952 1732 M 1952 1752 L
1932 1752 L
1932 1732 L
1952 1732 L
CF M
2081 1736 M B
2091 1726 M 2091 1746 L
2071 1746 L
2071 1726 L
2091 1726 L
CF M
2219 1731 M B
2229 1721 M 2229 1741 L
2209 1741 L
2209 1721 L
2229 1721 L
CF M
282 1967 M 421 1921 L
559 1885 L
697 1857 L
836 1834 L
974 1815 L
1112 1799 L
1251 1785 L
1389 1774 L
1527 1764 L
1666 1756 L
1804 1748 L
1942 1742 L
2081 1736 L
2219 1731 L
282 1849 M B
292 1839 M 292 1859 L
273 1859 L
273 1839 L
292 1839 L
CF M
421 1813 M B
431 1803 M 431 1823 L
411 1823 L
411 1803 L
431 1803 L
CF M
559 1786 M B
569 1776 M 569 1796 L
549 1796 L
549 1776 L
569 1776 L
CF M
697 1764 M B
707 1754 M 707 1774 L
687 1774 L
687 1754 L
707 1754 L
CF M
836 1747 M B
846 1737 M 846 1757 L
826 1757 L
826 1737 L
846 1737 L
CF M
974 1733 M B
984 1723 M 984 1743 L
964 1743 L
964 1723 L
984 1723 L
CF M
1112 1722 M B
1122 1712 M 1122 1731 L
1102 1731 L
1102 1712 L
1122 1712 L
CF M
1251 1712 M B
1261 1702 M 1261 1722 L
1241 1722 L
1241 1702 L
1261 1702 L
CF M
1389 1704 M B
1399 1694 M 1399 1714 L
1379 1714 L
1379 1694 L
1399 1694 L
CF M
1527 1697 M B
1537 1688 M 1537 1707 L
1517 1707 L
1517 1688 L
1537 1688 L
CF M
1666 1692 M B
1676 1682 M 1676 1702 L
1656 1702 L
1656 1682 L
1676 1682 L
CF M
1804 1687 M B
1814 1677 M 1814 1697 L
1794 1697 L
1794 1677 L
1814 1677 L
CF M
1942 1683 M B
1952 1673 M 1952 1693 L
1932 1693 L
1932 1673 L
1952 1673 L
CF M
2081 1680 M B
2091 1670 M 2091 1690 L
2071 1690 L
2071 1670 L
2091 1670 L
CF M
2219 1677 M B
2229 1667 M 2229 1687 L
2209 1687 L
2209 1667 L
2229 1667 L
CF M
282 1849 M 421 1813 L
559 1786 L
697 1764 L
836 1747 L
974 1733 L
1112 1722 L
1251 1712 L
1389 1704 L
1527 1697 L
1666 1692 L
1804 1687 L
1942 1683 L
2081 1680 L
2219 1677 L
282 1722 M B
292 1712 M 292 1732 L
273 1732 L
273 1712 L
292 1712 L
CF M
421 1697 M B
431 1688 M 431 1707 L
411 1707 L
411 1688 L
431 1688 L
CF M
559 1679 M B
569 1669 M 569 1689 L
549 1689 L
549 1669 L
569 1669 L
CF M
697 1665 M B
707 1655 M 707 1674 L
687 1674 L
687 1655 L
707 1655 L
CF M
836 1654 M B
846 1644 M 846 1664 L
826 1664 L
826 1644 L
846 1644 L
CF M
974 1645 M B
984 1635 M 984 1655 L
964 1655 L
964 1635 L
984 1635 L
CF M
1112 1638 M B
1122 1628 M 1122 1648 L
1102 1648 L
1102 1628 L
1122 1628 L
CF M
1251 1633 M B
1261 1623 M 1261 1643 L
1241 1643 L
1241 1623 L
1261 1623 L
CF M
1389 1629 M B
1399 1619 M 1399 1639 L
1379 1639 L
1379 1619 L
1399 1619 L
CF M
1527 1626 M B
1537 1616 M 1537 1636 L
1517 1636 L
1517 1616 L
1537 1616 L
CF M
1666 1623 M B
1676 1613 M 1676 1633 L
1656 1633 L
1656 1613 L
1676 1613 L
CF M
1804 1621 M B
1814 1611 M 1814 1631 L
1794 1631 L
1794 1611 L
1814 1611 L
CF M
1942 1620 M B
1952 1610 M 1952 1629 L
1932 1629 L
1932 1610 L
1952 1610 L
CF M
2081 1618 M B
2091 1609 M 2091 1628 L
2071 1628 L
2071 1609 L
2091 1609 L
CF M
2219 1618 M B
2229 1608 M 2229 1628 L
2209 1628 L
2209 1608 L
2229 1608 L
CF M
282 1722 M 421 1697 L
559 1679 L
697 1665 L
836 1654 L
974 1645 L
1112 1638 L
1251 1633 L
1389 1629 L
1527 1626 L
1666 1623 L
1804 1621 L
1942 1620 L
2081 1618 L
2219 1618 L
282 1583 M B
292 1573 M 292 1593 L
273 1593 L
273 1573 L
292 1573 L
CF M
421 1569 M B
431 1559 M 431 1579 L
411 1579 L
411 1559 L
431 1559 L
CF M
559 1560 M B
569 1550 M 569 1570 L
549 1570 L
549 1550 L
569 1550 L
CF M
697 1553 M B
707 1544 M 707 1563 L
687 1563 L
687 1544 L
707 1544 L
CF M
836 1549 M B
846 1539 M 846 1559 L
826 1559 L
826 1539 L
846 1539 L
CF M
974 1547 M B
984 1537 M 984 1557 L
964 1557 L
964 1537 L
984 1537 L
CF M
1112 1545 M B
1122 1535 M 1122 1555 L
1102 1555 L
1102 1535 L
1122 1535 L
CF M
1251 1544 M B
1261 1535 M 1261 1554 L
1241 1554 L
1241 1535 L
1261 1535 L
CF M
1389 1544 M B
1399 1535 M 1399 1554 L
1379 1554 L
1379 1535 L
1399 1535 L
CF M
1527 1545 M B
1537 1535 M 1537 1555 L
1517 1555 L
1517 1535 L
1537 1535 L
CF M
1666 1546 M B
1676 1536 M 1676 1556 L
1656 1556 L
1656 1536 L
1676 1536 L
CF M
1804 1547 M B
1814 1537 M 1814 1557 L
1794 1557 L
1794 1537 L
1814 1537 L
CF M
1942 1548 M B
1952 1538 M 1952 1558 L
1932 1558 L
1932 1538 L
1952 1538 L
CF M
2081 1550 M B
2091 1540 M 2091 1560 L
2071 1560 L
2071 1540 L
2091 1540 L
CF M
2219 1552 M B
2229 1542 M 2229 1562 L
2209 1562 L
2209 1542 L
2229 1542 L
CF M
282 1583 M 421 1569 L
559 1560 L
697 1553 L
836 1549 L
974 1547 L
1112 1545 L
1251 1544 L
1389 1544 L
1527 1545 L
1666 1546 L
1804 1547 L
1942 1548 L
2081 1550 L
2219 1552 L
282 1415 M B
292 1405 M 292 1425 L
273 1425 L
273 1405 L
292 1405 L
CF M
421 1414 M B
431 1404 M 431 1424 L
411 1424 L
411 1404 L
431 1404 L
CF M
559 1416 M B
569 1406 M 569 1426 L
549 1426 L
549 1406 L
569 1406 L
CF M
697 1419 M B
707 1409 M 707 1429 L
687 1429 L
687 1409 L
707 1409 L
CF M
836 1423 M B
846 1413 M 846 1433 L
826 1433 L
826 1413 L
846 1413 L
CF M
974 1427 M B
984 1417 M 984 1437 L
964 1437 L
964 1417 L
984 1417 L
CF M
1112 1432 M B
1122 1422 M 1122 1442 L
1102 1442 L
1102 1422 L
1122 1422 L
CF M
1251 1437 M B
1261 1427 M 1261 1447 L
1241 1447 L
1241 1427 L
1261 1427 L
CF M
1389 1442 M B
1399 1432 M 1399 1452 L
1379 1452 L
1379 1432 L
1399 1432 L
CF M
1527 1447 M B
1537 1437 M 1537 1457 L
1517 1457 L
1517 1437 L
1537 1437 L
CF M
1666 1452 M B
1676 1442 M 1676 1462 L
1656 1462 L
1656 1442 L
1676 1442 L
CF M
1804 1457 M B
1814 1447 M 1814 1467 L
1794 1467 L
1794 1447 L
1814 1447 L
CF M
1942 1462 M B
1952 1452 M 1952 1472 L
1932 1472 L
1932 1452 L
1952 1452 L
CF M
2081 1467 M B
2091 1457 M 2091 1477 L
2071 1477 L
2071 1457 L
2091 1457 L
CF M
2219 1472 M B
2229 1462 M 2229 1482 L
2209 1482 L
2209 1462 L
2229 1462 L
CF M
282 1415 M 421 1414 L
559 1416 L
697 1419 L
836 1423 L
974 1427 L
1112 1432 L
1251 1437 L
1389 1442 L
1527 1447 L
1666 1452 L
1804 1457 L
1942 1462 L
2081 1467 L
2219 1472 L
282 1059 M B
292 1049 M 292 1069 L
273 1069 L
273 1049 L
292 1049 L
CF M
421 1085 M B
431 1076 M 431 1095 L
411 1095 L
411 1076 L
431 1076 L
CF M
559 1110 M B
569 1100 M 569 1120 L
549 1120 L
549 1100 L
569 1100 L
CF M
697 1133 M B
707 1123 M 707 1143 L
687 1143 L
687 1123 L
707 1123 L
CF M
836 1155 M B
846 1145 M 846 1165 L
826 1165 L
826 1145 L
846 1145 L
CF M
974 1175 M B
984 1165 M 984 1185 L
964 1185 L
964 1165 L
984 1165 L
CF M
1112 1194 M B
1122 1184 M 1122 1204 L
1102 1204 L
1102 1184 L
1122 1184 L
CF M
1251 1211 M B
1261 1202 M 1261 1221 L
1241 1221 L
1241 1202 L
1261 1202 L
CF M
1389 1228 M B
1399 1218 M 1399 1238 L
1379 1238 L
1379 1218 L
1399 1218 L
CF M
1527 1244 M B
1537 1234 M 1537 1254 L
1517 1254 L
1517 1234 L
1537 1234 L
CF M
1666 1259 M B
1676 1249 M 1676 1269 L
1656 1269 L
1656 1249 L
1676 1249 L
CF M
1804 1273 M B
1814 1263 M 1814 1283 L
1794 1283 L
1794 1263 L
1814 1263 L
CF M
1942 1286 M B
1952 1276 M 1952 1296 L
1932 1296 L
1932 1276 L
1952 1276 L
CF M
2081 1298 M B
2091 1288 M 2091 1308 L
2071 1308 L
2071 1288 L
2091 1288 L
CF M
2219 1310 M B
2229 1300 M 2229 1320 L
2209 1320 L
2209 1300 L
2229 1300 L
CF M
282 1059 M 421 1085 L
559 1110 L
697 1133 L
836 1155 L
974 1175 L
1112 1194 L
1251 1211 L
1389 1228 L
1527 1244 L
1666 1259 L
1804 1273 L
1942 1286 L
2081 1298 L
2219 1310 L
CS [] 0 setdash M
1105 100 M 1126 151 M 1126 100 L
1129 151 M 1129 100 L
1112 151 M 1110 136 L
1110 151 L
1146 151 L
1146 136 L
1143 151 L
1119 100 M 1136 100 L
1216 160 M 1211 156 L
1206 148 L
1201 139 L
1199 127 L
1199 117 L
1201 105 L
1206 95 L
1211 88 L
1216 83 L
1211 156 M 1206 146 L
1203 139 L
1201 127 L
1201 117 L
1203 105 L
1206 98 L
1211 88 L
1235 151 M 1235 100 L
1237 151 M 1252 108 L
1235 151 M 1252 100 L
1268 151 M 1252 100 L
1268 151 M 1268 100 L
1271 151 M 1271 100 L
1228 151 M 1237 151 L
1268 151 M 1278 151 L
1228 100 M 1242 100 L
1261 100 M 1278 100 L
1293 120 M 1321 120 L
1321 124 L
1319 129 L
1317 132 L
1312 134 L
1305 134 L
1297 132 L
1293 127 L
1290 120 L
1290 115 L
1293 108 L
1297 103 L
1305 100 L
1309 100 L
1317 103 L
1321 108 L
1319 120 M 1319 127 L
1317 132 L
1305 134 M 1300 132 L
1295 127 L
1293 120 L
1293 115 L
1295 108 L
1300 103 L
1305 100 L
1336 151 M 1353 100 L
1338 151 M 1353 108 L
1370 151 M 1353 100 L
1331 151 M 1345 151 L
1360 151 M 1374 151 L
1384 160 M 1389 156 L
1394 148 L
1398 139 L
1401 127 L
1401 117 L
1398 105 L
1394 95 L
1389 88 L
1384 83 L
1389 156 M 1394 146 L
1396 139 L
1398 127 L
1398 117 L
1396 105 L
1394 98 L
1389 88 L
CS [] 0 setdash M
CS [] 0 setdash M
146 1188 M 117 1219 M 112 1222 L
122 1222 L
117 1219 L
115 1217 L
112 1212 L
112 1202 L
115 1197 L
117 1195 L
122 1195 L
124 1197 L
127 1202 L
132 1214 L
134 1219 L
136 1222 L
120 1195 M 122 1197 L
124 1202 L
129 1214 L
132 1219 L
134 1222 L
141 1222 L
144 1219 L
146 1214 L
146 1205 L
144 1200 L
141 1197 L
136 1195 L
146 1195 L
141 1197 L
86 1277 M 163 1234 L
134 1291 M 141 1294 L
144 1296 L
146 1301 L
146 1306 L
144 1313 L
136 1318 L
129 1320 L
122 1320 L
117 1318 L
115 1315 L
112 1311 L
112 1306 L
115 1299 L
122 1294 L
129 1291 L
163 1282 L
146 1306 M 144 1311 L
136 1315 L
129 1318 L
120 1318 L
115 1315 L
112 1306 M 115 1301 L
122 1296 L
129 1294 L
163 1284 L
CS [] 0 setdash M
1527 1940 M CS [] 0 setdash M
1361 1914 M 1378 1948 M 1364 1897 L
1381 1948 M 1366 1897 L
1378 1941 M 1376 1926 L
1376 1919 L
1381 1914 L
1385 1914 L
1390 1917 L
1395 1921 L
1400 1929 L
1405 1948 M 1397 1921 L
1397 1917 L
1400 1914 L
1407 1914 L
1412 1919 L
1414 1924 L
1407 1948 M 1400 1921 L
1400 1917 L
1402 1914 L
1426 1943 M 1470 1943 L
1426 1929 M 1470 1929 L
1501 1965 M 1494 1962 L
1489 1955 L
1486 1943 L
1486 1936 L
1489 1924 L
1494 1917 L
1501 1914 L
1506 1914 L
1513 1917 L
1518 1924 L
1520 1936 L
1520 1943 L
1518 1955 L
1513 1962 L
1506 1965 L
1501 1965 L
1496 1962 L
1494 1960 L
1491 1955 L
1489 1943 L
1489 1936 L
1491 1924 L
1494 1919 L
1496 1917 L
1501 1914 L
1506 1914 M 1511 1917 L
1513 1919 L
1515 1924 L
1518 1936 L
1518 1943 L
1515 1955 L
1513 1960 L
1511 1962 L
1506 1965 L
CS [] 0 setdash M
1527 1839 M CS [] 0 setdash M
1431 1814 M 1441 1855 M 1443 1853 L
1441 1850 L
1438 1853 L
1438 1855 L
1441 1860 L
1443 1862 L
1450 1865 L
1460 1865 L
1467 1862 L
1470 1860 L
1472 1855 L
1472 1850 L
1470 1845 L
1462 1841 L
1450 1836 L
1446 1833 L
1441 1828 L
1438 1821 L
1438 1814 L
1460 1865 M 1465 1862 L
1467 1860 L
1470 1855 L
1470 1850 L
1467 1845 L
1460 1841 L
1450 1836 L
1438 1819 M 1441 1821 L
1446 1821 L
1458 1816 L
1465 1816 L
1470 1819 L
1472 1821 L
1446 1821 M 1458 1814 L
1467 1814 L
1470 1816 L
1472 1821 L
1472 1826 L
1501 1865 M 1494 1862 L
1489 1855 L
1486 1843 L
1486 1836 L
1489 1824 L
1494 1816 L
1501 1814 L
1506 1814 L
1513 1816 L
1518 1824 L
1520 1836 L
1520 1843 L
1518 1855 L
1513 1862 L
1506 1865 L
1501 1865 L
1496 1862 L
1494 1860 L
1491 1855 L
1489 1843 L
1489 1836 L
1491 1824 L
1494 1819 L
1496 1816 L
1501 1814 L
1506 1814 M 1511 1816 L
1513 1819 L
1515 1824 L
1518 1836 L
1518 1843 L
1515 1855 L
1513 1860 L
1511 1862 L
1506 1865 L
CS [] 0 setdash M
1527 1779 M CS [] 0 setdash M
1431 1754 M 1460 1800 M 1460 1754 L
1462 1804 M 1462 1754 L
1462 1804 M 1436 1768 L
1474 1768 L
1453 1754 M 1470 1754 L
1501 1804 M 1494 1802 L
1489 1795 L
1486 1783 L
1486 1775 L
1489 1763 L
1494 1756 L
1501 1754 L
1506 1754 L
1513 1756 L
1518 1763 L
1520 1775 L
1520 1783 L
1518 1795 L
1513 1802 L
1506 1804 L
1501 1804 L
1496 1802 L
1494 1800 L
1491 1795 L
1489 1783 L
1489 1775 L
1491 1763 L
CS M
1494 1759 L
1496 1756 L
1501 1754 L
1506 1754 M 1511 1756 L
1513 1759 L
1515 1763 L
1518 1775 L
1518 1783 L
1515 1795 L
1513 1800 L
1511 1802 L
1506 1804 L
CS [] 0 setdash M
1527 1719 M CS [] 0 setdash M
1431 1694 M 1467 1737 M 1465 1735 L
1467 1732 L
1470 1735 L
1470 1737 L
1467 1742 L
1462 1744 L
1455 1744 L
1448 1742 L
1443 1737 L
1441 1732 L
1438 1722 L
1438 1708 L
1441 1701 L
1446 1696 L
1453 1694 L
1458 1694 L
1465 1696 L
1470 1701 L
1472 1708 L
1472 1710 L
1470 1718 L
1465 1722 L
1458 1725 L
1455 1725 L
1448 1722 L
1443 1718 L
1441 1710 L
1455 1744 M 1450 1742 L
1446 1737 L
1443 1732 L
1441 1722 L
1441 1708 L
1443 1701 L
1448 1696 L
1453 1694 L
1458 1694 M 1462 1696 L
1467 1701 L
1470 1708 L
1470 1710 L
1467 1718 L
1462 1722 L
1458 1725 L
1501 1744 M 1494 1742 L
1489 1735 L
1486 1722 L
1486 1715 L
1489 1703 L
1494 1696 L
1501 1694 L
1506 1694 L
1513 1696 L
1518 1703 L
1520 1715 L
1520 1722 L
1518 1735 L
1513 1742 L
1506 1744 L
1501 1744 L
1496 1742 L
1494 1739 L
1491 1735 L
1489 1722 L
1489 1715 L
1491 1703 L
1494 1698 L
1496 1696 L
1501 1694 L
1506 1694 M 1511 1696 L
1513 1698 L
1515 1703 L
1518 1715 L
1518 1722 L
1515 1735 L
1513 1739 L
1511 1742 L
1506 1744 L
CS [] 0 setdash M
1527 1659 M CS [] 0 setdash M
1431 1634 M 1450 1684 M 1443 1682 L
1441 1677 L
1441 1670 L
1443 1665 L
1450 1662 L
1460 1662 L
1467 1665 L
1470 1670 L
1470 1677 L
1467 1682 L
1460 1684 L
1450 1684 L
1446 1682 L
1443 1677 L
1443 1670 L
1446 1665 L
1450 1662 L
1460 1662 M 1465 1665 L
1467 1670 L
1467 1677 L
1465 1682 L
1460 1684 L
1450 1662 M 1443 1660 L
1441 1658 L
1438 1653 L
1438 1643 L
1441 1638 L
1443 1636 L
1450 1634 L
1460 1634 L
1467 1636 L
1470 1638 L
1472 1643 L
1472 1653 L
1470 1658 L
1467 1660 L
1460 1662 L
1450 1662 M 1446 1660 L
1443 1658 L
1441 1653 L
1441 1643 L
1443 1638 L
1446 1636 L
1450 1634 L
1460 1634 M 1465 1636 L
1467 1638 L
1470 1643 L
1470 1653 L
1467 1658 L
1465 1660 L
1460 1662 L
1501 1684 M 1494 1682 L
1489 1674 L
1486 1662 L
1486 1655 L
1489 1643 L
1494 1636 L
1501 1634 L
1506 1634 L
1513 1636 L
1518 1643 L
1520 1655 L
1520 1662 L
1518 1674 L
1513 1682 L
1506 1684 L
1501 1684 L
1496 1682 L
1494 1679 L
1491 1674 L
1489 1662 L
1489 1655 L
1491 1643 L
1494 1638 L
1496 1636 L
1501 1634 L
1506 1634 M 1511 1636 L
1513 1638 L
1515 1643 L
1518 1655 L
1518 1662 L
1515 1674 L
1513 1679 L
1511 1682 L
1506 1684 L
CS [] 0 setdash M
1527 1579 M CS [] 0 setdash M
1383 1553 M 1397 1594 M 1402 1597 L
1409 1604 L
1409 1553 L
1407 1601 M 1407 1553 L
1397 1553 M 1419 1553 L
1453 1604 M 1446 1601 L
1441 1594 L
1438 1582 L
1438 1575 L
1441 1563 L
1446 1556 L
1453 1553 L
1458 1553 L
1465 1556 L
1470 1563 L
1472 1575 L
1472 1582 L
1470 1594 L
1465 1601 L
1458 1604 L
1453 1604 L
1448 1601 L
1446 1599 L
1443 1594 L
1441 1582 L
1441 1575 L
1443 1563 L
1446 1558 L
1448 1556 L
1453 1553 L
1458 1553 M 1462 1556 L
1465 1558 L
1467 1563 L
1470 1575 L
1470 1582 L
1467 1594 L
1465 1599 L
1462 1601 L
1458 1604 L
1501 1604 M 1494 1601 L
1489 1594 L
1486 1582 L
1486 1575 L
1489 1563 L
1494 1556 L
1501 1553 L
1506 1553 L
1513 1556 L
1518 1563 L
1520 1575 L
1520 1582 L
1518 1594 L
1513 1601 L
1506 1604 L
1501 1604 L
1496 1601 L
1494 1599 L
1491 1594 L
1489 1582 L
1489 1575 L
1491 1563 L
1494 1558 L
1496 1556 L
1501 1553 L
1506 1553 M 1511 1556 L
1513 1558 L
1515 1563 L
1518 1575 L
1518 1582 L
1515 1594 L
1513 1599 L
1511 1601 L
1506 1604 L
CS [] 0 setdash M
1527 1478 M CS [] 0 setdash M
1383 1453 M 1397 1494 M 1402 1496 L
1409 1503 L
1409 1453 L
1407 1501 M 1407 1453 L
1397 1453 M 1419 1453 L
1441 1494 M 1443 1491 L
1441 1489 L
1438 1491 L
1438 1494 L
1441 1499 L
1443 1501 L
1450 1503 L
1460 1503 L
1467 1501 L
1470 1499 L
1472 1494 L
1472 1489 L
1470 1484 L
1462 1479 L
1450 1475 L
1446 1472 L
1441 1467 L
1438 1460 L
1438 1453 L
1460 1503 M 1465 1501 L
1467 1499 L
1470 1494 L
1470 1489 L
1467 1484 L
1460 1479 L
1450 1475 L
1438 1458 M 1441 1460 L
1446 1460 L
1458 1455 L
1465 1455 L
1470 1458 L
1472 1460 L
1446 1460 M 1458 1453 L
1467 1453 L
1470 1455 L
1472 1460 L
1472 1465 L
1501 1503 M 1494 1501 L
1489 1494 L
1486 1482 L
1486 1475 L
1489 1463 L
1494 1455 L
1501 1453 L
1506 1453 L
1513 1455 L
1518 1463 L
1520 1475 L
1520 1482 L
CS M
1518 1494 L
1513 1501 L
1506 1503 L
1501 1503 L
1496 1501 L
1494 1499 L
1491 1494 L
1489 1482 L
1489 1475 L
1491 1463 L
1494 1458 L
1496 1455 L
1501 1453 L
1506 1453 M 1511 1455 L
1513 1458 L
1515 1463 L
1518 1475 L
1518 1482 L
1515 1494 L
1513 1499 L
1511 1501 L
1506 1503 L
CS [] 0 setdash M
1527 1298 M CS [] 0 setdash M
1383 1272 M 1397 1313 M 1402 1316 L
1409 1323 L
1409 1272 L
1407 1321 M 1407 1272 L
1397 1272 M 1419 1272 L
1460 1318 M 1460 1272 L
1462 1323 M 1462 1272 L
1462 1323 M 1436 1287 L
1474 1287 L
1453 1272 M 1470 1272 L
1501 1323 M 1494 1321 L
1489 1313 L
1486 1301 L
1486 1294 L
1489 1282 L
1494 1275 L
1501 1272 L
1506 1272 L
1513 1275 L
1518 1282 L
1520 1294 L
1520 1301 L
1518 1313 L
1513 1321 L
1506 1323 L
1501 1323 L
1496 1321 L
1494 1318 L
1491 1313 L
1489 1301 L
1489 1294 L
1491 1282 L
1494 1277 L
1496 1275 L
1501 1272 L
1506 1272 M 1511 1275 L
1513 1277 L
1515 1282 L
1518 1294 L
1518 1301 L
1515 1313 L
1513 1318 L
1511 1321 L
1506 1323 L
CS [] 0 setdash M
stroke
grestore
showpage
end
{\footnotesize
\begin{thebibliography}{99}
%
\bibitem{Ra92}
K. Rajagopal and F. Wilczek,
`Static and Dynamic Critical Phenomena at a Second Order QCD Phase Transition',
Princeton preprint PUPT-1347 (1992)
\bibitem{La80} C. M. Lattes, Y. Fujimoto and S. Hasegawa,
Phys. Rep. {\bf 65}, 151 (1990)
\bibitem{Ko92}
K. L. Kowalski and C. C. Taylor,
`Disoriented Chiral Condensates: A white paper for the Full Acceptance
Detector', Case Western Reserve University preprint 92-6 (1992)
\bibitem{Ho71}
D. Horn and R. Silver, Ann. Phys. {\bf 66}, 509 (1971)
\bibitem{Gy79}
M. Gyulassy, S. K. Kauffmann and L. W. Wilson,
Phys. Rev. C{\bf 20}, 2267 (1979)
\bibitem{Ka90}
M. Kataja and P. V. Ruuskanen, Phys. Lett. B{\bf 243}, 181 (1990)
\bibitem{Ge90}
P. Gerber, H. Leutwyler and J. L. Goity, Phys. Lett. B{\bf 246}, 513 (1990)
\bibitem{Mu92}
K. Geiger and J. I. Kapusta, Phys. Rev. D{\bf 47}, 4905 (1993);
\\
E. V. Shuryak, Phys. Rev. Lett. {\bf 68}, 3270 (1992);
\\
B. M\"uller, `Physics of the Quark Gluon Plasma', Duke preprint TH-92-36
(1992);
\\
T. S. Bir\'{o}, E. van Doorn, B. M\"uller, M. H. Thoma and X.N. Wang,
`Parton equilibration in relativistic heavy ion collisions',
Duke preprint TH-93-46 (1993)
\bibitem{Sh91}
E. V. Shuryak, Nucl. Phys. A{\bf 533}, 761 (1991);
\\
V. Koch and G. F. Bertsch, Nucl. Phys. A{\bf 552}, 591 (1993)
\bibitem{We92}
G. Welke and G. F. Bertsch, Phys. Rev. C{\bf 45}, 1403 (1992)
\bibitem{Be92}
H. Bebie, P. Gerber, J. L. Goity and H. Leutwyler,
Nucl. Phys. B{\bf 378}, 95 (1992)
\bibitem{Be88}
G. Bertsch, M. Gong, L. McLerran, V. Ruuskanen and E. Sarkkinen,
Phys. Rev. D{\bf 37}, 1202 (1988)
\bibitem{Pr92}
S. Pratt, Phys. Lett. B{\bf 301}, 159 (1993)
\bibitem{Lam}
C. S. Lam and S. Y. Lo, Phys. Rev. Lett. {\bf 52}, 1184 (1984)
%
%

\end{thebibliography}
}
